%% file: paper_v0.tex
\documentclass[pre,twocolumn,aps,10pt,superscriptaddress,notitlepage,nofootinbib]{revtex4-1}  

\usepackage{amsmath}
\usepackage{bm}
\usepackage{graphicx}
\usepackage{placeins}

\usepackage[usenames,dvipsnames,svgnames,table]{xcolor}
\usepackage{hyperref}
\hypersetup{
	colorlinks   = true,
	citecolor    = blue
}
\usepackage{cleveref}

\crefname{equation}{Eq.}{Eqs.}
\crefname{figure}{Fig.}{Figs.}
\crefname{appendix}{Appendix}{Appendices1}

\graphicspath{{./Fig/}}

\DeclareMathOperator{\sign}{sgn}

\newcommand{\hi}{\hat{i}}
\newcommand{\hj}{\hat{j}}

\newcommand{\us}{\underline{\sigma} }
\newcommand{\bs}{\boldsymbol{\sigma}}
\newcommand{\ubs}{\underline{\bs} }

\newcommand{\Parenthesis}[1]{\left( #1 \right)}

\newcommand{\Avr}[1]{\overline{ #1} }

\newcommand{\Graph}{\mathcal{G}}
\newcommand{\Neighbor}[1]{ {\partial #1} }
\newcommand{\NeighborExcept}[2]{ {\partial #1 \setminus #2} }
\newcommand{\LocalConstraint}[1]{\psi_{#1}(\underline{\sigma}_{#1},  \ubs_{\Neighbor{#1} } )}
\newcommand{\Message}[2]{m_{#1 \to #2} (\us_{#1}, \us_{#2})}
\RequirePackage[normalem]{ulem} 
\RequirePackage{color}\definecolor{RED}{rgb}{1,0,0}\definecolor{BLUE}{rgb}{0,0,1} 
\providecommand{\DIFaddbegin}{} 
\providecommand{\DIFaddend}{} 
\providecommand{\DIFdelbegin}{} 
\providecommand{\DIFdelend}{} 
\providecommand{\DIFaddbeginFL}{} 
\providecommand{\DIFaddendFL}{} 
\providecommand{\DIFdelbeginFL}{} 
\providecommand{\DIFdelendFL}{} 
\newcommand{\DIFscaledelfig}{0.5}
\RequirePackage{settobox} 
\RequirePackage{letltxmacro} 
\newsavebox{\DIFdelgraphicsbox} 
\newlength{\DIFdelgraphicswidth} 
\newlength{\DIFdelgraphicsheight} 
\LetLtxMacro{\DIFOincludegraphics}{\includegraphics} 
\newcommand{\DIFaddincludegraphics}[2][]{{\color{blue}\fbox{\DIFOincludegraphics[#1]{#2}}}} 
\newcommand{\DIFdelincludegraphics}[2][]{
\sbox{\DIFdelgraphicsbox}{\DIFOincludegraphics[#1]{#2}}
\settoboxwidth{\DIFdelgraphicswidth}{\DIFdelgraphicsbox} 
\settoboxtotalheight{\DIFdelgraphicsheight}{\DIFdelgraphicsbox} 
\scalebox{\DIFscaledelfig}{
\parbox[b]{\DIFdelgraphicswidth}{\usebox{\DIFdelgraphicsbox}\\[-\baselineskip] \rule{\DIFdelgraphicswidth}{0em}}\llap{\resizebox{\DIFdelgraphicswidth}{\DIFdelgraphicsheight}{
\setlength{\unitlength}{\DIFdelgraphicswidth}
\begin{picture}(1,1)
\thicklines\linethickness{2pt} 
{\color[rgb]{1,0,0}\put(0,0){\framebox(1,1){}}}
{\color[rgb]{1,0,0}\put(0,0){\line( 1,1){1}}}
{\color[rgb]{1,0,0}\put(0,1){\line(1,-1){1}}}
\end{picture}
}\hspace*{3pt}}} 
} 
\LetLtxMacro{\DIFOaddbegin}{\DIFaddbegin} 
\LetLtxMacro{\DIFOaddend}{\DIFaddend} 
\LetLtxMacro{\DIFOdelbegin}{\DIFdelbegin} 
\LetLtxMacro{\DIFOdelend}{\DIFdelend} 
\DeclareRobustCommand{\DIFaddbegin}{\DIFOaddbegin \let\includegraphics\DIFaddincludegraphics} 
\DeclareRobustCommand{\DIFaddend}{\DIFOaddend \let\includegraphics\DIFOincludegraphics} 
\DeclareRobustCommand{\DIFdelbegin}{\DIFOdelbegin \let\includegraphics\DIFdelincludegraphics} 
\DeclareRobustCommand{\DIFdelend}{\DIFOaddend \let\includegraphics\DIFOincludegraphics} 
\LetLtxMacro{\DIFOaddbeginFL}{\DIFaddbeginFL} 
\LetLtxMacro{\DIFOaddendFL}{\DIFaddendFL} 
\LetLtxMacro{\DIFOdelbeginFL}{\DIFdelbeginFL} 
\LetLtxMacro{\DIFOdelendFL}{\DIFdelendFL} 
\DeclareRobustCommand{\DIFaddbeginFL}{\DIFOaddbeginFL \let\includegraphics\DIFaddincludegraphics} 
\DeclareRobustCommand{\DIFaddendFL}{\DIFOaddendFL \let\includegraphics\DIFOincludegraphics} 
\DeclareRobustCommand{\DIFdelbeginFL}{\DIFOdelbeginFL \let\includegraphics\DIFdelincludegraphics} 
\DeclareRobustCommand{\DIFdelendFL}{\DIFOaddendFL \let\includegraphics\DIFOincludegraphics} 

\begin{document}

\title{On the number of limit cycles in diluted neural networks.}

\author{Sungmin Hwang}
\affiliation{LPTMS, Universit\'e Paris-Sud 11, UMR 8626 CNRS, B\^at. 100, 91405 Orsay Cedex, France}
\author{Enrico Lanza}
\affiliation{Dipartimento di Biotecnologie Cellulari ed Ematologia, Sapienza Universit\`a
	di Roma, P.le A. Moro 2, 00185 Roma, Italy}
\author{Giorgio Parisi}
\affiliation{Dipartimento di Fisica, Sapienza Universit\`a di Roma, P.le A. Moro 2, 00185
	Roma, Italy}
\author{Jacopo Rocchi}
\affiliation{LPTMS, Universit\'e Paris-Sud 11, UMR 8626 CNRS, B\^at. 100, 91405 Orsay Cedex, France}
\author{Giancarlo Ruocco}
\affiliation{Center for Life Nano Science, Fondazione Istituto Italiano di Tecnologia (IIT), Viale Regina Elena 291, I00161 Roma, Italy}
\author{Francesco Zamponi}
\affiliation{Laboratoire de Physique de l'Ecole Normale Sup\'erieure, ENS, Universit\'e PSL, CNRS, Sorbonne Universit\'e, Universit\'e de Paris, F-75005 Paris, France}

\date{\today}

\begin{abstract}
We consider the storage properties of temporal patterns, i.e. cycles of finite lengths, in neural networks represented by (generally asymmetric) spin glasses defined on random graphs. Inspired by the observation that dynamics on sparse systems has more basins of attractions than the dynamics of densely connected ones, we consider the attractors of a greedy dynamics in sparse topologies, considered as proxy for the stored memories. We enumerate them using numerical simulation and extend the analysis to large systems sizes using belief propagation. We find that the logarithm of the number of such cycles is a non monotonic function of the mean connectivity and we discuss the similarities with biological neural networks describing the memory capacity of the hippocampus. 

\end{abstract}

\maketitle

\section{Introduction}

\textcolor{red}{}

Identifying the relation between single neuron activity and the mechanisms underlying the cognitive processes and collective properties of complexly interconnected networks is one of the most important questions in neuroscience. 
In spite of the highly organized structure, the mammalian brain is impressively complex, with billions of neurons and ten thousand times as many connections supporting the integration of information from different areas. 
Facing such a high complexity, we are forced to resort to simplified models to dissect the problem in clear and controllable elements. 
Artificial neural networks offer a way to model and study nervous systems with arbitrarily simple connectivity. To this extent, understanding the dynamics of these models in the resulting information storage and processing at the level of network topology is a way to investigate crucial elements affecting the behaviour of complex nervous system. 

The present work takes up where Ref.~\cite{hwang2019number} left off, further investigating the relationship between connectivity structure properties of recurrent networks and their storing power, associated with the resulting number of limit cycles, which represent cyclically repeating patterns of neuronal activity.
As content-addressable model for such a Hopfield network \cite{Hopfield1982}, we employ a recurrent network of McCulloch-Pitts neurons \cite{McCulloch1943}, which presents deterministic dynamics.
In recurrent neural networks, information storage is based on the association of input patterns to cyclic activity (limit cycles, or attractors), which encode memorized elements. Considering the deterministic dynamics, this association is necessarily unequivocal, making it a reliable way to store information non-locally. 
The retrieval time of a memory encoded in such a way, depends on the strength of the associated neuronal patterns. 
The connectivity structure of a system like this completely determines the clustering operation mapping the input set into the generally smaller set of attractors.

The identification of information storage with cyclical activation patterns, as the ones featured in Hopfield models, proved to have experimental support in resting state recordings of neuronal populations in the hippocampus of rodents \cite{Pfeiffer2015} and in task driven neuronal activity in monkeys \cite{Fuster1971,Miyashita1988}.
Additionally, the Hopfield model found many benefits from the striking analogy it bears with spin systems, which belong to a category of complex systems that has been extensively studied in physics, from its introduction \cite{Heisenberg1928} until today \cite{Amit1985, amit1985storing, amit1987statistical, amit1992modeling}. Both models represent networks of relatively simple elementary units, whose dynamics result from the interaction among elements. The ``neighboring'' elements are completely defined by the connectivity matrix $J$, playing a key role in such a system.
In the case of neuronal networks, the connectivity matrix contains the information about the \textit{connectome}, i.e. the set of synaptic connections of either chemical or electrical nature. 
The construction of the connectome may follow two different paths. On one side, it can be based on synaptic plasticity, which means that a preexisting `empty' network is enriched with new connections every time a new memory is produced. 
Alternatively, when storing memories, preexisting connections of an already filled matrix may be modified (reinforcement). In any case, memories are represented by attractors of the dynamics defined by the preexisting random network and the maximum storage capacity is linearly dependent on the system size ($\approx0.14N$).
This is the framework we focus on, studying the properties of random $J$ matrices. 

Tanaka and Edwards \cite{Tanaka1980} identify the relation between the mean number of fixed points and the system size of random ensembles of fully connected Ising spin glass models at thermodynamic limit. In this case, the binary variables are spins $\sigma_{i} \in\{-1,1\}$ and again, the dynamics are driven by the strength of the interaction of neighboring nodes, which are defined by asymmetric matrices $J$, whose $J_{ij}$ elements are drawn from a fixed distribution with zero mean.
By including a symmetry parameter (shown to influence the number of limit cycles as well \cite{Gutfreund1988}), another study focusing on this kind of model at finite size show the presence of a transition point linked to the symmetry of the connectivity matrix \cite{Bastolla1998}. Systems with higher symmetry feature fixed points or limit cycles of length 2, whose number increase exponentially with the system size. Systems with lower symmetry parameter present limit cycles whose length increases exponentially with the system size. Other transitions obtained with this model regard the transient time spent to reach the attractor from random inputs \cite{Gutfreund1988,Bastolla1998,Nutzel1991} and the size of the basins of attraction of the existing limit cycles \cite{Bastolla1998}. Adding noise \cite{Molgedey1992,schuecker2018optimal}, dilution \cite{Tirozzi1991}, a gain function \cite{Sompolinsky1988,Crisanti1990,crisanti2018path} or self-interaction \cite{Stern2014} may further expand the array of possible transitions in the system. In particular, adding dilution to fully asymmetric networks results in an increased number of attractors, and therefore to an increased storage capacity \cite{folli2018effect}.

Our previous work \cite{hwang2019number} focused on the mean number of limit cycles of arbitrary length $L$ and their associated complexity $\Sigma_{L}$. In particular, we presented a method based on \cite{GardnerE.1987} that enables to compute, for fully connected matrices and different symmetry degrees, the average number of limit cycles of any length in the form $n_{L}\sim (A^{L}/L)\, \text{exp}(N\Sigma_L)$, and solved the numerical computation of the average for lengths $L\leq3$. We also included information on cycle states overlap, which is related to the attractor structure, and showed the presence of a transition to chaotic regimes, depending on the symmetry degree of the network.

Here, we focus on the complexity of limit cycles of length 4 at the thermodynamic limit, in diluted matrices as a function of the connectivity parameter $c$ and the symmetry parameter $\epsilon$.
The problem of storing temporal patterns in a supervised manner has been considered in \cite{baldassi2013theory}. Here we consider a similar problem in an unsupervised context. We argue that for fully asymmetric networks, these are the smallest cycle lengths allowed, and that longer cycles have lengths that are multiples of four. Additionally, we investigate the finite size effect of the model with three different ensembles of random matrices and show how the $\Sigma_4$ complexity at finite sizes converges towards the fully connected case as $c$ is increased.

\section{Model}
\label{sec:method}

We consider a system of $N$ binary neurons, encoded a vector spin variable $\bs$ with entries ${\sigma_i \in \{-1,1 \}}$ 
($i=1,\dots,N$).
At each discrete time step $t$, $\bs^t$ undergoes a deterministic dynamics which updates all nodes synchronously according to the following rule:
\begin{equation}
\label{Dynamics}
\sigma_i^{t+1}=\sign \Parenthesis{
	\sum\limits_{j=1}^N J_{ij}\sigma_j^{t}
},
\end{equation}
where $\sign(x)$ is the sign function. \DIFdelbegin 

\DIFdelend The couplings $J_{ij}$'s are quenched random variables which stay constant over the dynamical process. To study the average properties of the dynamics, we introduce matrix ensembles for $J_{ij}$ constructed as follows.
First, regarding topology, we consider two commonly used graph ensembles that control the overall connectivity of networks, namely, regular random graphs and Erd\H{o}s-R\'enyi (ER) graphs. 
For both cases, as long as the average connectivity $c$, or the mean degree, is kept constant in the limit $N \to \infty$, the resulting graphs are locally tree-like and their adjacency matrices are sparse. 

Next, for each graph realization $\Graph$, we assign the coupling constants for connected links considering bidirectional couplings such that both $J_{ij}$ and $J_{ji}$ are nonzero if $i$ and $j$ are connected, following experimental evidence commonly found in real cortical networks. 
We further assume that the coupling $J_{ij}$ and $J_{ji}$ are constructed from two random numbers of the form:
\begin{equation}
\label{Coupling}
J_{ij}=\left(1-\frac{\epsilon}{2}\right)S_{ij}+\frac{\epsilon}{2}A_{ij},
\end{equation}
where $S_{ij}$ and $A_{ij}$ are i.i.d. standard Gaussian random variables with the symmetric and antisymmetric conditions, i.e., $S_{ji}$=$S_{ij}$, $A_{ji}$=$-A_{ij}$, respectively. 
The parameter $\epsilon$ controls the overall asymmetry of coupling constants, the special cases $\epsilon = (0, 1, 2)$ corresponding to 
symmetric, asymmetric and antisymmetric coupling matrices, respectively.

Here, we mainly focus on the long-time properties of the dynamics, i.e., the statistical properties of periodic points (or limit cycles) of the dynamics. 
In the following, we present a general formalism that computes the average number of $L$-cycles in the limit $N\rightarrow\infty$ 
by extending the pioneering work by Gardner, Derrida and Mottishaw~\cite{GardnerE.1987}. 

Given two spin configurations $\bs, \bs'$, we construct an indicator variable $w(\bs, \bs')$, which yields a binary value equal to one if
$\bs'$ is the one-step evolution of $\bs$ according to \cref{Dynamics}, and zero otherwise.
This indicator variable is conveniently represented by observing that 
$\bs'$ is the evolution of $\bs$ if and only if the
local field $H_i(\bs_\Neighbor{i}) = \sum_{j \in \Neighbor{i} } J_{ij} \sigma_j$ has the same sign as $\sigma'_i$, for all $i$.
Here, the symbol $ \bs_\Neighbor{i} $ is introduced to denote the set of neighboring spins of $i$, which emphasizes the fact that $ H_i(\bs_\Neighbor{i}) $ only depends on the spin configuration of the neighbors of node $i$.
The above condition allows us to encode $ w (\bs,\bs') $  in a compact way as a product of Heaviside theta functions:
\begin{align} 
\label{Indicator}
w (\bs,\bs') = {\prod _{i=1}^{N}} 
	\theta \left( \sigma'_i H_i(\bs_{\Neighbor{i}}) \right).
\end{align}
By applying this logic repeatedly, it is clear that the product 
$ \prod_{t=1}^{L} w (\bs^t,\bs^{t+1})$ 
is again an indicator variable that detects the trajectory $\ubs = \{ \bs^1 , \cdots, \bs^L, \bs^{L+1} \}$.
Here and from now on, the underline symbol denotes a trajectory $\ubs = \{ \bs^1 , \cdots, \bs^L, \bs^{L+1} \}$ and $\us_i = \{ \sigma_i^1 , \cdots, \sigma_i^L, \sigma_i^{L+1} \}$ denotes its single entry.
With the additional periodic boundary condition $ \bs^{L+1} = \bs^1 $, the trajectory is a cycle of length $L$.
Specifically, we define the partition function:
\begin{align}
Z_L = \sum_{ \ubs } \prod_{t=1}^{L} w (\bs^t,\bs^{t+1})
\equiv \sum_{ \ubs } {\prod _{i=1}^{N}} \LocalConstraint{i},
\label{PartitionFunctionDef}
\end{align}
where the summation over $\ubs$ contains all possible spin trajectories obeying the periodic boundary condition. 
In the last equality, we rearrange the terms to represent the partition function as the product of local constraints 
\begin{align}
\LocalConstraint{i} = \prod_{t=1}^{L} \theta \left( \sigma^{t+1}_i H_i(\bs^t_{\Neighbor{i}}) \right).
\label{LocalConstraint}
\end{align}
By explicitly writing out their arguments, we stress that each local constraint $ \LocalConstraint{i} $ only depends on the configurations of the node $i$ and of its neighbors.

One caveat follows from the formalism.
This periodic boundary condition in \cref{PartitionFunctionDef} is not only satisfied by $L$-cycles but also by cycles of length $L'$ if $L'$ is a divisor of $L$, i.e., $L' | L$.
Thus, denoting the number of $L$-cycles by $n_L$, the following identity holds
\begin{equation}
Z_L = \sum_{L' | L}  n_{L'} L' \ ,
\label{PartitionFuncNCycleRelation}
\end{equation}
where the additional factor $L$ comes from the fact that each $L$-cycle consists of $L$ distinct spin configurations.
Thus to compute $n_L$ exactly, one needs to evaluate $Z_{L'}$ for all divisors $L'$ and reconstruct $n_{L'}$ recursively from \cref{PartitionFuncNCycleRelation}.
This procedure corresponds to M\"obius inversion formula of a Dirichlet convolution.

We use a powerful technique, the so-called \textit{cavity equation} or \textit{belief propagation equations} (BP) \cite{MPV87,MP01}, 
which efficiently computes $Z_L$ in a linear time in $N$ in contrast to an exponential time complexity exhibited by the naive brute-force counting.
This method is guaranteed to work if two conditions are met: 
	i) the underlying graph structure is locally tree-like and 
	ii) the partition function is given as the product of local constraints $ \LocalConstraint{i} $ summed over the configurations.
In the problem under consideration, the usual BP framework needs to be generalised \cite{yedidia2001characterization, yedidia2001generalized} and it can be used once the problems has been set on a dual graph representation, explained in \cref{appendix:BP} and recently used in \cite{lokhov2015dynamic, rocchi2018slow}.
Under these conditions, the partition function is decomposed into two contributions: i) $z_i$ coming from the nodes $i$ and 
ii) $z_{(ij)}$ from the links $(ij)$ on the underlying graph $\Graph$, respectively. 
Namely, we have
\begin{align}
\log Z_{L} = \sum_{i} \log z_{i} - \sum_{(ij)} \log z_{(ij)},
\label{PartitionFunctionFinal}
\end{align}
where $z_{i} $ and $z_{(ij)}$ are defined below.
The derivation of such decomposition relies on a super-factor representation and is detailed in \cref{appendix:BP}. \DIFdelbegin 

\DIFdelend Furthermore, one can show that each contribution is solely determined by a single set of \emph{messages} $\Message{i}{j}$ in the following way
\begin{equation}
z_{(ij)} = \sum_{\sigma_i, \sigma_j} \Message{i}{j} \Message{j}{i}
\label{ZLink}
\end{equation}
and
\begin{equation}
z_i = \sum_{\sigma_i} \sum_{\us_{\partial i}} \LocalConstraint{i} \prod_{k \in \partial i } \Message{k}{i},
\label{ZNode}
\end{equation}
where the messages satisfy a recursive relation
\begin{equation}
\Message{i}{j} \propto \sum_{\ubs_\NeighborExcept{i}{j}} \LocalConstraint{i} \prod_{k \in \NeighborExcept{i}{j}} \Message{k}{i}.
\label{Messageitoj}
\end{equation}
With the additional normalization condition $ \sum_{\us_i, \us_j} \Message{i}{j} =1 $,
this set of recursive equations can be solved iteratively. 
Starting from an arbitrarily chosen set of messages, the right hand side of \cref{Messageitoj} is computed repeatedly until convergence.
This iterative scheme is guaranteed to work on a tree-like factor graph, which is the case for our problem in the super-factor representation as explained in \cref{appendix:BP}. 

\section{Results}
With the powerful theoretical tool for computing \cref{PartitionFuncNCycleRelation} at hand, we are ready to compute the ensemble average over many samples up to a sufficiently large system size $N \sim 1000$.
Specifically, we focus on computing the quenched average $\Avr{\ln Z_L}$, which is a self-averaging quantity as opposed to the annealed average $\Avr{Z_L}$.
Mainly, we want to estimate the exponential growth coefficient $\Sigma_L$ by extracting the linear part from the usual $ 1/N $ expansion:
\begin{align}
	\Avr{\ln Z_L} \sim \Sigma_L N +B_L + O(N^{-1}).
	\label{PartitionFunctionQuenched}
\end{align}

We specifically focus on the case $L=4$ because cycles of smaller length are not allowed in the infinite $N$ limit. As discussed in \mbox{
\cref{appendix:Motif}}\hspace{0pt}
, this is due to the existence of a graph motif which only forms periodic orbits of length $L=4$. The probability of finding these motifs in a graph can be arbitrarily small depending on the model parameters $c$ or $\epsilon>0$, but the probability per node is nonzero and independent of $N$, implying that they will eventually appear if the system size is taken to be large enough. Therefore, the smallest possible cycles in the system are of length $L=4$ and cycles should be of length equal to a multiple of $4$. This statement is confirmed by our numerical findings. Strictly speaking, one cannot exclude the possibility of finding other motifs that further prevent the existence of cycles of length $L=4$. 
However, both our brute force counting and our results from the cavity method suggest that $4$-cycles always yield the dominant contributions to the partition function and thus at least within the system sizes we consider, $N \leq 1000$, we may conclude that either such motifs do not exist or they are rare enough to not be found within our numerical procedures. 
On the other hand, solving our BP equations takes a time growing exponentially in $L$, as a result of the increasing dimension of the state variables, due
to the length of trajectories increasing linearly in $L$, see \cref{Messageitoj}. Computing $\Sigma_L$ for the next non-trivial length $L=8$ is thus computationally very challenging.
Unless specified otherwise, in the following we focus on the special case $\epsilon=1$.

\subsection{Thermodynamic limit of the random regular graph}

\begin{figure}
	\includegraphics[width=0.8\linewidth]{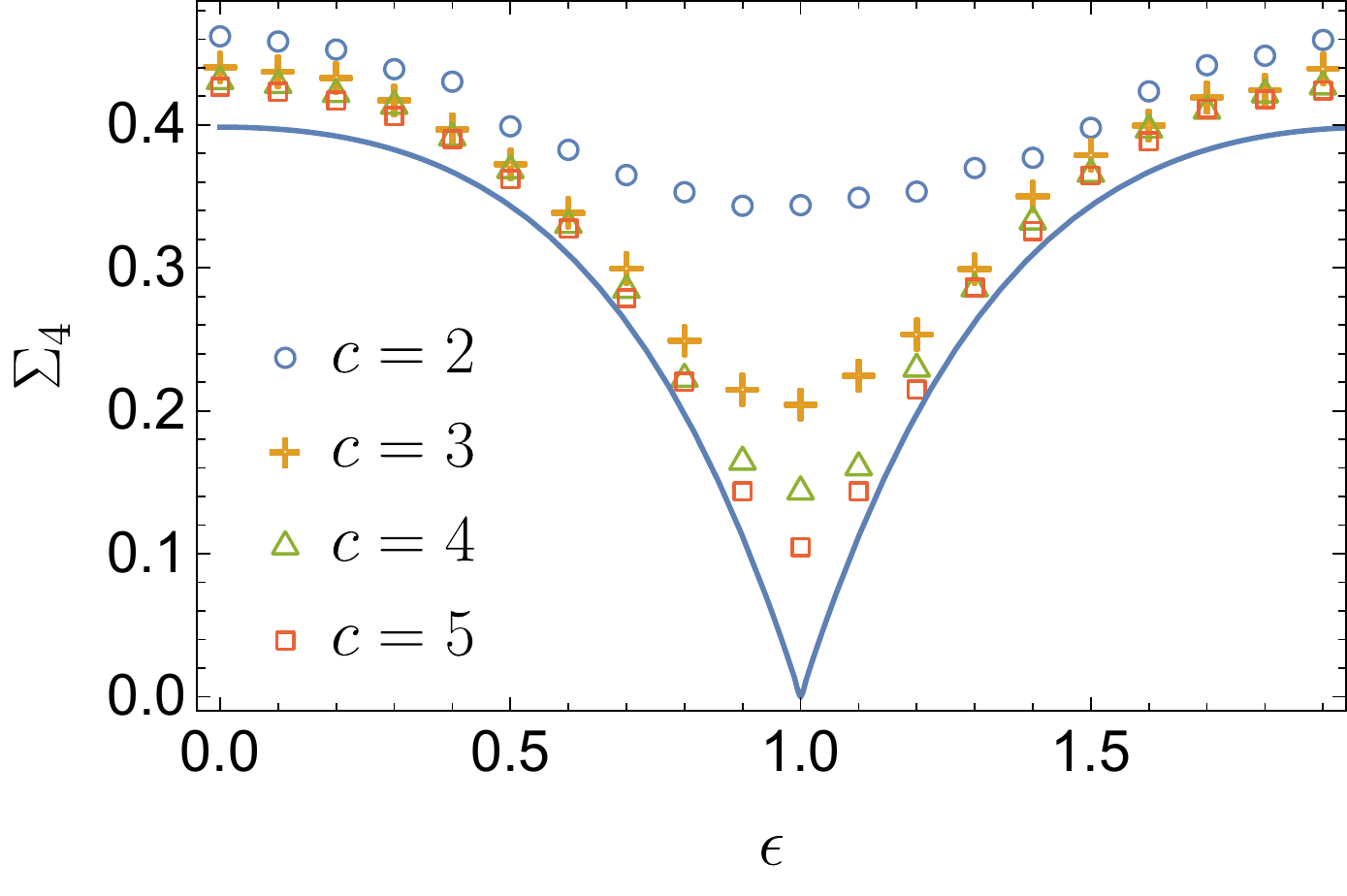}
	\caption{Quenched complexity $\Sigma_4$ as a function of $\epsilon$ for various $c$, extrapolated to the thermodynamic limit.
		The complexity is seen to be symmetric around the line $\epsilon =1$ at which the lowest value of $\Sigma_4$ is achieved. 
		The solid line indicates the \emph{annealed} complexity $\Sigma_4^{\mathrm{ann}}$ for the fully-connected model~\cite{hwang2019number}. 
		Quenched and annealed averages are not guaranteed to be the same. 
		Nevertheless, as $c$ increases, the quenched complexity (slowly) converges to the annealed one, see~\cref{fig:Sigma4Epsilon1}. 
	}
	\label{fig:Sigma4}
\end{figure}

\begin{figure}
\includegraphics[width=.8\linewidth]{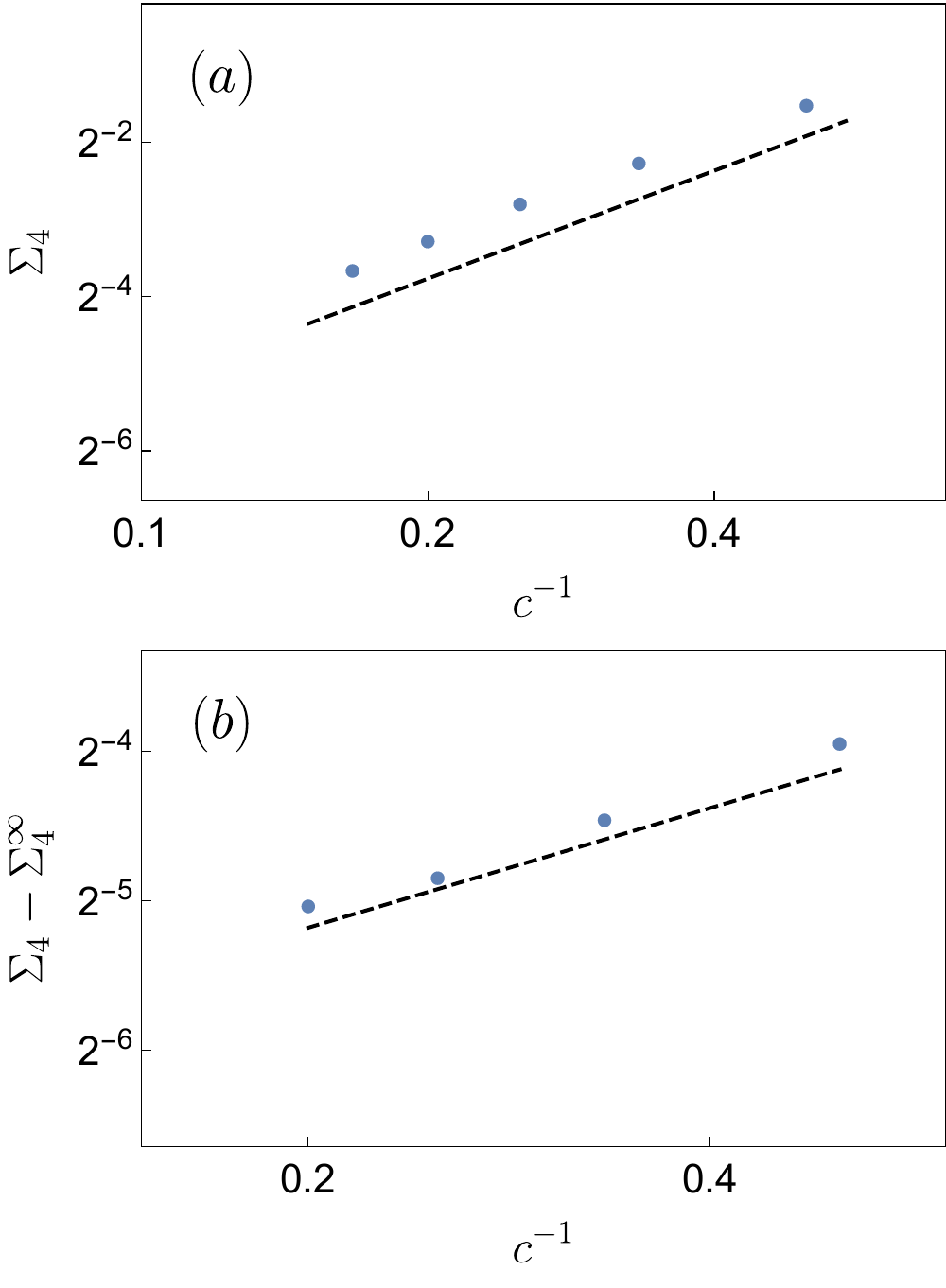}
\caption{Quenched complexity $\Sigma_4$, extrapolated to the thermodynamic limit, as a function of $1/c$ at $\epsilon =1$ (a) and $\epsilon=0$ (b). 	
		As expected from the fully-connected case, the complexity decreases monotonously with increasing $c$.
		The decay is reasonably described by a power-law of the form (a) $\Sigma_4 \sim c^{-1.4}$ and (b) $\Sigma_4 - \Sigma_4^{\infty} \sim c^{-0.8}$ respectively.
		The decay exponent for (b) is significantly slower than the one for (a). 
	}
	\label{fig:Sigma4Epsilon1}
\end{figure}

Now, let us discuss our numerical findings.
First of all, we present the results on the random regular (RR) graph ensemble in which the connectivity is fixed to be $c$, thus no fluctuations are associated to the degree distribution. When dealing with different ensembles of graphs, we generally use $c$ to denote mean connectivity, except in the RR case where it refers to the fixed node degree. In \cref{fig:Sigma4}, we present the quenched complexity $\Sigma_4(c)$, extrapolated to the thermodynamic limit using~\cref{PartitionFunctionQuenched} with data for $N\leq 1000$, 
and different connectivity $c = 2,3,4,5$, as well as the annealed complexity 
$\Sigma_4^{\mathrm{ann}}(c\to\infty)$, which is computed analytically for the fully connected graphs 
in the thermodynamic limit~\cite{hwang2019number}. The complexity attains its minimum at $\epsilon = 1$, i.e. for fully asymmetric couplings.
In fact, since $J_{ij}$ and $J_{ji}$ are completely independent, 
the formation of cycles should rely purely on the random occurrences.
Moreover, at a fixed $\epsilon$, we clearly see the decreasing behavior of $\Sigma_4$ as a function of $c$.  

Note that $\Sigma_4$ may converge or not to the fully connected annealed result (solid line) in the limit $c\to\infty$, because the quenched complexity is not necessarily the same as the annealed one. On the other hand, the Jensen's inequality implies that for any $c,N$,
\begin{equation}
\Sigma_4(c,N) \leq \Sigma_4^{\mathrm{ann}}(c,N) \ .
\end{equation}
Hence, the limit for $c\to\infty$
of $\Sigma_4(c,N)$ should be lesser or equal than the fully connected annealed result. We attribute the observation 
that in \cref{fig:Sigma4} the data seem to converge to a slightly higher
value than the solid line to a slow convergence with $c$, as we now discuss.

\begin{figure}
	\centering
    \includegraphics[width=.5\textwidth]{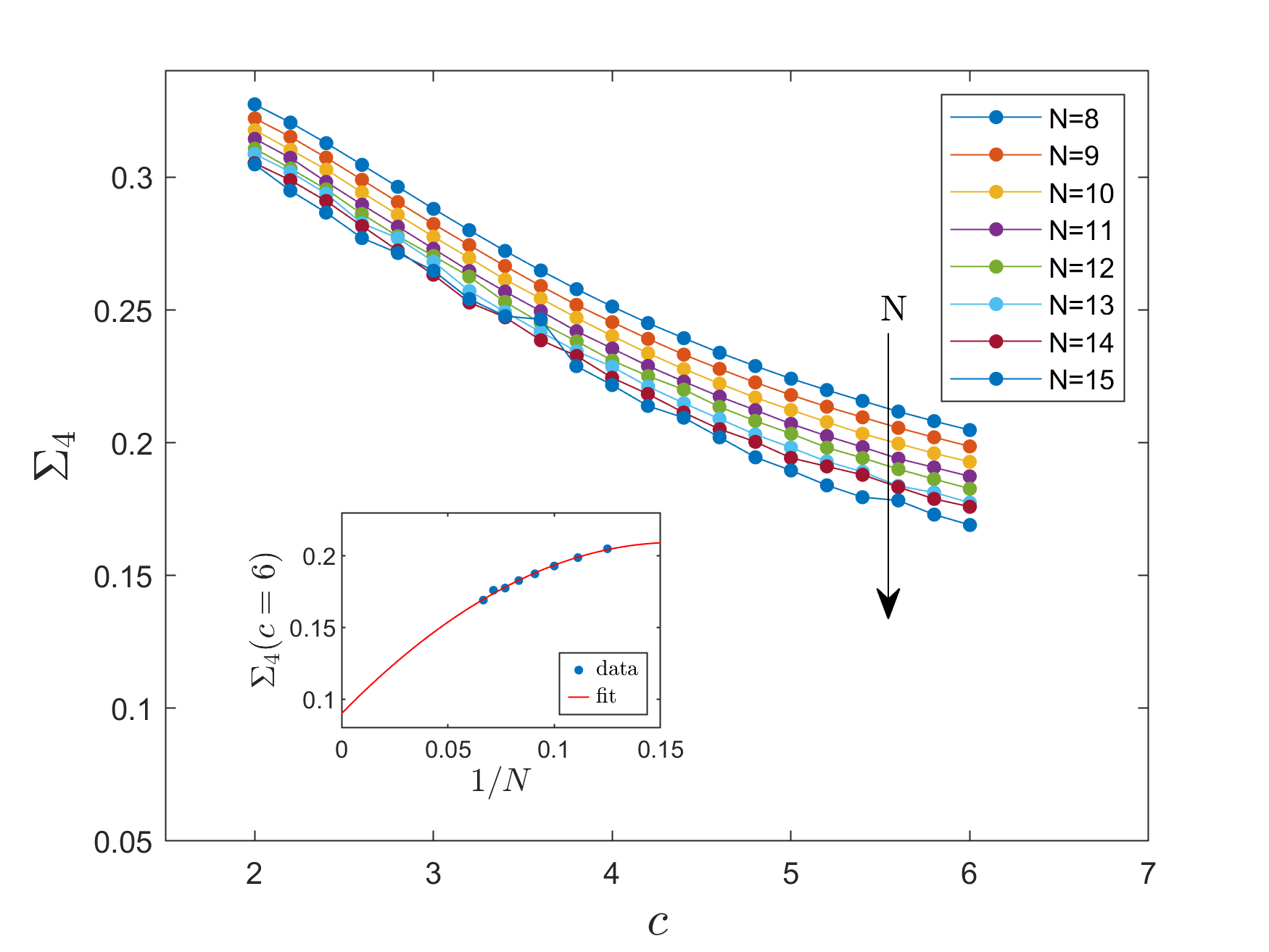}
	\caption{Quenched complexity $\Sigma_4$ as a function of $c$ at $\epsilon =1$ for finite size systems with $N$ ranging from 8 to 15. (Inset) Quenched complexity $\Sigma_4$ at $c=6$ as a function of $1/N$ at $\epsilon =1$, together with a quadratic fit $\Sigma_4(N) = \Sigma_4(\infty) + a/N + b/N^2$, with 
	$\Sigma_4(\infty)=0.09$ (taken from~\cref{fig:Sigma4Epsilon1}), $a=1.515$, $b=-4.812$. 
	The consistency between the data points and the fit support the results reported in Fig. \ref{fig:Sigma4Epsilon1}.
	}
	\label{fig:Sigma4Epsilon1FiniteSize}
\end{figure}

In \cref{fig:Sigma4Epsilon1} we present the behavior of $\Sigma_4(c)$, extrapolated to $N\to\infty$ as above, 
as a function of $c$ for $\epsilon=1$ and $\epsilon=0$.
Because (i) $\Sigma_4^{\mathrm{ann}}(\epsilon=1) =0$ for the fully connected system, (ii) the annealed complexity is an upper bound of the quenched one, 
and (iii) the complexity cannot be negative,
we should expect $\Sigma_4$ to converge to zero as $c$ increases.
Consistently with this argument, in \cref{fig:Sigma4Epsilon1}, 
we show that $\Sigma_4$ decays as a power-law of the form $\Sigma_4 \sim c^{-\alpha}$ with an exponent $\alpha \approx 1.4$. 
For $\epsilon=0$, our data show that $\Sigma_4$ also converges to the annealed result for $c\to\infty$, this time with an exponent $\alpha\approx0.8$.

\begin{figure}
	\centering
    \includegraphics[width=.79\linewidth]{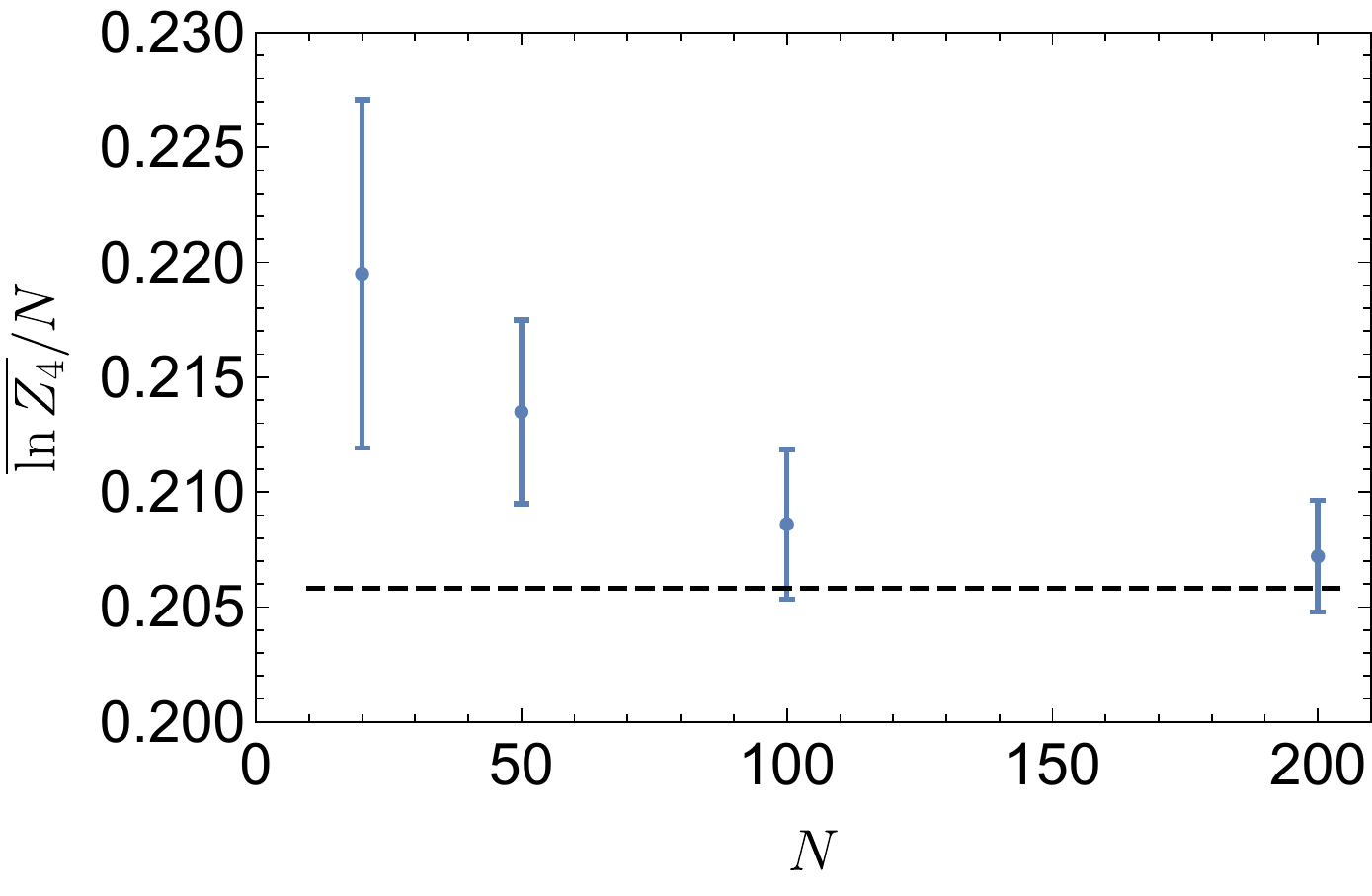}
	\caption{Plot of complexity  $\Sigma_4(N)=\Avr{\ln Z_4}/N$ vs $N$ for $c=4$.
	It shows the rate of decay of the finite-size correction to $\Sigma_4$. 
	The dashed line indicates the estimate of $\Sigma_4$ from the ansatz \cref{PartitionFunctionQuenched}.	
	The error bars shows the sample-to-sample fluctuations for different system sizes.
	The variability decreases monotonously with increasing $N$ implying that $\Sigma_4$ is self-averaging. 
	}
	\label{fig:FiniteCorrection}
\end{figure}

\subsection{Finite size corrections}

Next, we discuss the finite size effects in our numerical results for $\Sigma_4(N) = \Avr{\ln Z_4}/N$.
The results for the quenched complexity obtained by exact enumeration from finite size systems with sizes ranging from $N=8$ to 15 are reported 
in~\cref{fig:Sigma4Epsilon1FiniteSize}.
 \cref{fig:FiniteCorrection} shows the same quantity but now obtained from the cavity method for much larger systems.  
As indicated by the $1/N$ expansion in \cref{PartitionFunctionQuenched}, we can extrapolate to the infinite size limit and we observe that the result obtained at $N=200$ is close enough to the asymptotic value. Besides, the sample-to-sample fluctuations for different system sizes are reported. 
Consistently with our theoretical prediction of $\Avr{\ln Z_4}/N$ being self-averaging, we find a decreasing variability with system size $N$.

\subsection{Other graph ensembles}

\begin{figure}[t]
\centerline{\resizebox{1.1\linewidth}{!}{\input{Fig/fig5.tex}}}
\caption{Complexity $\Sigma_4$ at $\epsilon=1$ for different ensembles and connectivity at $N=100$. The DP label refers to the ensemble discussed in the main text. Inset: size of the largest component in the graph. Results are averaged over $100$ random realizations for each ensemble.}
\label{fig:fig-Comp-Ens}
\end{figure}
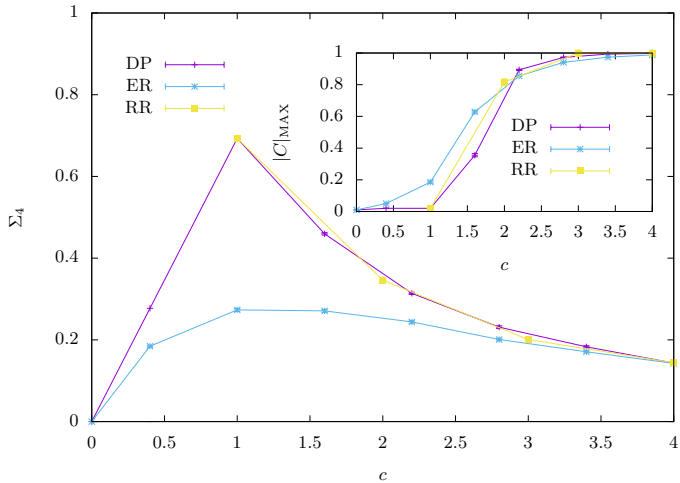

We compute the complexity as a function of the average connectivity in different ensembles of graphs, see Fig. \ref{fig:fig-Comp-Ens}.
For ER graphs, we observe that the complexity is a non monotonic function of the mean connectivity. 
On the other hand, for RR graphs, we observe that it is a decreasing function of the mean connectivity. 
A possible interpretation of this phenomenon is that, once a giant connected cluster exists, each time a connection is established, some cycle disappears.
The first increase in ER graphs at low connectivity is due to a very sparse regime with many isolated nodes and without a giant connected cluster.
To test this hypothesis, we consider a third ensemble of graphs, built according to the following rule. Be $M$ the number of total links. For a graph of $N$ sites, we can form $N/2$ different links before allowing any node to have degree larger than $2$. When all these pairings have been made, each node is exactly connected to another node, leading to $c=1$. Only after having paired any node with another random one, random nodes can be paired by the creation of other links. We denote this ensemble by the acronym DP (Dyadic Pair).
In this ensemble, adding new links at low connectivity leads to a linear increase of the complexity. This effect is smoothed in ER graphs because, at the same value of $c$ with respect to a DP graph, some of the links are added between already connected nodes, leading to a smaller increase of the connectivity with $c$.

The behavior observed at low connectivity depends on the choice to exclude isolated nodes (i.e. nodes with degree equal to zero) contributions from the computation of the complexity. In fact, the case of isolated nodes seems pathological at first sight because, having no neighbors, they are not considered in the dynamical update defined in Eq.~(\ref{Dynamics}). One possible choice to extend the dynamical rule to cover this case would be to set the state of $\sigma_i^t$ equal to $+1$ or $-1$ with probability $1/2$. In this case, the dynamics would be completely symmetric with respect to the up-down symmetry. The non-monotonous behavior of the complexity as a function of the connectivity would disappear in this case, leading in particular to the result $\Sigma_4=\log 2$ at $c=0$, because any configuration would trivially satisfy the constraints. 
On the other hand, there are no reasons for the dynamical update to be symmetric and the choice made here is equivalent to consider  $\sigma_i^t$ equal to $-1$  (or $+1$, or unchanged) for isolated nodes. In fact, allowing $\sigma_i^t$ to take one state only, we notice that their contribution to the number of cycles is just a factor one. Since the complexity of disconnected parts of the graph is the sum of their relative complexities, the contribution due to isolated nodes is zero.
Moreover, $\sigma_i^t$ equal to $-1$ bears a more physical content: if we consider spins to represent neurons, and $-1$ to represent the rest state, it makes more sense to consider neurons that remain at rest when no stimuli come to perturb them than neurons that randomly flip their state. 
Let us also notice that the number of such nodes at a given $c$ depends on the ensemble considered and, for $c<1$, ER graph contains more such nodes than a DP graph, explaining the lower growth of the complexity in the first case.

\section{Conclusion}
In Ref.~\cite{hwang2019number} we extended the analysis of \cite{Tanaka1980} deriving an analytical expression of the average number of limit cycles of length $L$, $\langle n^{L} \rangle$, as a function of the symmetry parameter $\epsilon$, \cref{Coupling}, in neural networks defined by \cref{Dynamics}, and solved it for $L\geq3$. 
 
 This work takes up where \cite{hwang2019number} left off. Starting from~\cref{PartitionFunctionQuenched}, we introduce here a new system parameter, namely the connectivity $c$, which is included as an additional argument of the complexity $\Sigma$, and is related to the degree of dilution of the network. We compute the quenched average 
 $\Sigma_L(N) = \overline{\text{ln}Z_{L}}/N$, reported in Fig. \ref{fig:Sigma4} as a function of $\epsilon$ for different values of connectivity $c$, together with the corresponding annealed average.

We argue that, at $\epsilon$=1, i.e. for fully asymmetric networks, the shortest limit cycles appearing in the dynamics are of length $L=4$, and that the length of longer cycles must be a multiple of $4$ ($L=4 n$, $n=1,2,3...$). We show that the hypothesis that these cycles provide the main contribution to the average of the total number of attractors is confirmed both by our brute force calculation and by the results of the cavity method. We also provide an explanation as to why there cannot be limit cycles with $L<4$ at thermodynamic limit, linking this effect to the occurrence of specific motifs preventing the appearance of shorter attractors.

Focusing on $\Sigma_{4}$, we report the complexity for $L=4$ cycles and we show that it is a decreasing function of the connectivity $c$ at $\epsilon$=1. In particular, the complexity may be approximated in the form $\Sigma_{4} \left(c,\epsilon=1\right) \sim  c^{-\alpha}$ , with $\alpha\approx 1.4$. Similar results are also obtained at $\epsilon$=0, but with a smaller
$\alpha$ exponent.
This analysis is enriched with the study of finite size effects, carried out on three different ensembles of random networks. The numerical results show how $\overline{\text{ln}Z_{L}}/N$ is close to the infinite size limit already at $N=200$. Additionally, its decreasing variability is in line with the hypothesis that this quantity is self-averaging. We also notice that adding new links at low connectivity ($ c  < 1$) increases the complexity, while doing so at high connectivity ($ c > 1$) results in the opposite effect, and associate this behavior with the creation of a giant cluster crucially altering the dynamics of the networks.

Taken together, these results are in line with previous findings, which prove that diluting random networks of large size results in a higher number of basins of attraction directly related to a higher storing capacity of the system \cite{folli2018effect}. Interestingly, diluted networks have also other important properties shared with biologic neural networks: sparse connections have been proved to reduce spiking related noise in cortical areas, contributing to enhance the reliability and the stability of the memory retrieval process \cite{rolls2012cortical}. 

Sparsity is a feature shared also by local areas of the hippocampus known as CA3: the probability of establishing a connection between two CA3 neurons is of 4\% in mammalian hippocampus and 10\% for pyramidal cells in neocortex \cite{rolls2012advantages,witter2010connectivity}.
To explain this observation, the authors of \cite{rolls2012advantages} show that increasing the connectivity reduces the memory capacity of the network.
This is consistent with the memory-oriented specialization of these brain areas, although other factors may contribute to define these particular structures, such as the retrieval time of the stored limit cycles. We notice that the `overloading' condition discussed in \cite{rolls2012advantages}, linked to a memory retrieval impairment due the shift of the energy landscape of the system towards fewer and bigger basins of attractions, shares the same flavour of the results presented in this study.

In conclusion, this work emphasizes the relationship between the observed structure of diluted brain areas and the crucial role of dilution in models of neuronal networks, which determines an optimization rule that may have shaped these brain areas to maximize the possible storage capacity.

\acknowledgments

We thank Guilhem Semerjian for useful discussions related to this work and in particular for contributing to the development of the procedure discussed
in \cref{appendix:BP}. J.R. and S.H. acknowledge the support of a grant from the Simons Foundation (No. 454941, Silvio Franz).

\begin{figure}
	\centering
	\includegraphics[width=\linewidth]{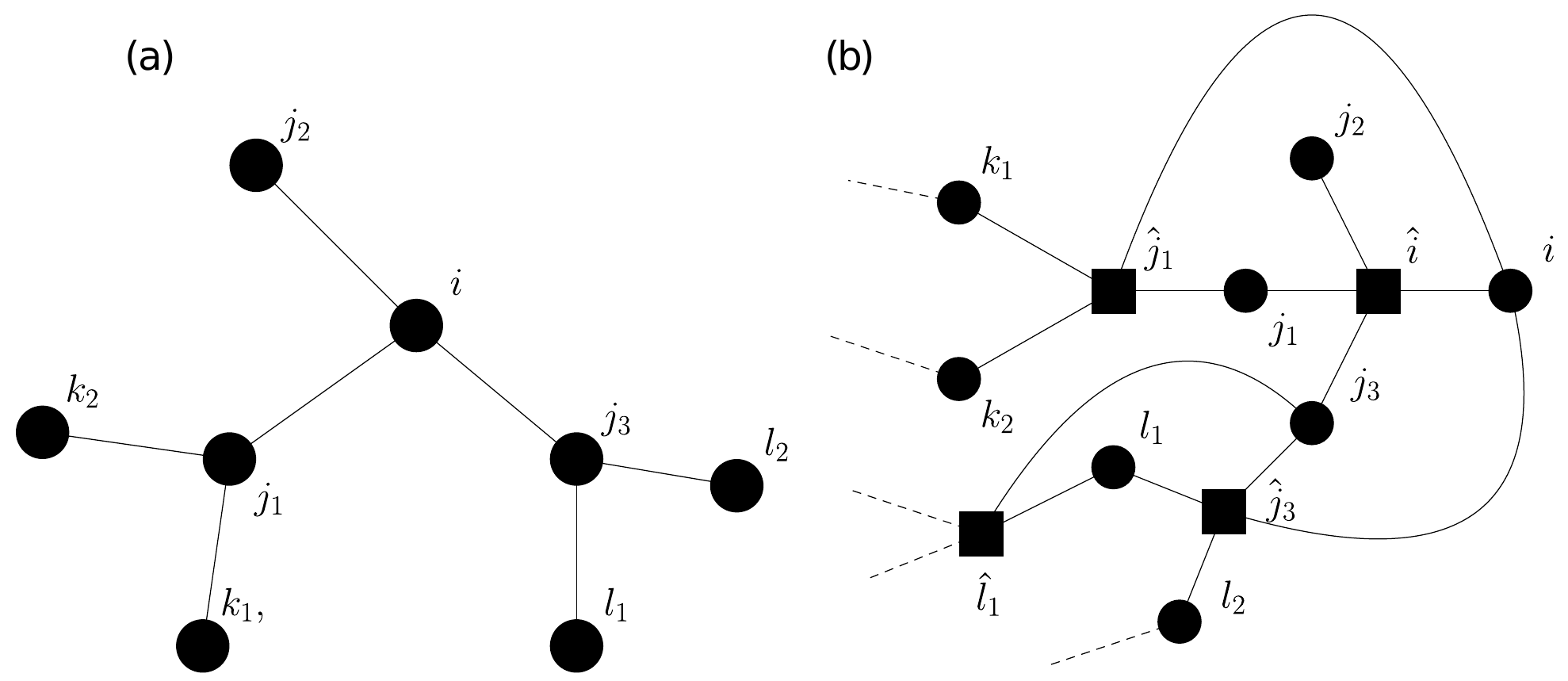}
	\caption{(a) Original graph topology $\Graph$ on which we consider a parallel dynamics. (b) Corresponding factor graph for the counting cycles problem. While $\mathcal{G}$ contains no loops, we see that the factor graph contains many short loops.}
	\label{fig:Graph}
\end{figure}

\begin{figure}
\centering
\includegraphics[width=1\linewidth]{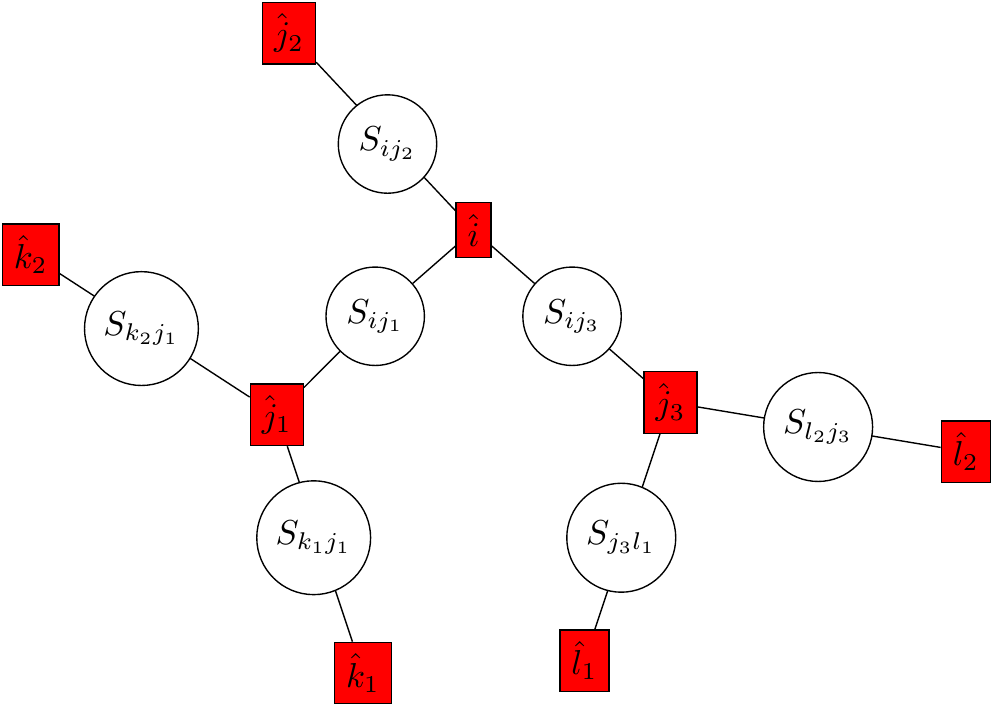}
\caption{
An equivalent super-factor graph representation of the original problem in \cref{fig:Graph} is shown.
For each node $i$ of the original graph $\mathcal{G}$ we assign a super-factor $\hi$ and for each link $(i,j)$ we assign a super-variable $S_{ij} = (\us_{ij}, \us_{ji})$.
At the cost of doubling the dimension of super variables, the resulting graph topology retains the same graph structure as $\Graph$, and thus it allows us to use a cavity method as long as the original graph $\Graph$ is tree-like.
The redundancy due to the increase of state dimension is then handled by implementing additional local constraints such that 
all the variables $\us_{ik}$ sharing the same first index are constrained to be the same.}
\label{fig:SuperFactor}
\end{figure}

\appendix

\section{BP equations}
\label{appendix:BP}
A well-known technique for computing the partition function \cref{PartitionFunctionDef} 
on a sparse graph is the so-called cavity method~\cite{MPV87,MP01}. 
Given the problem under consideration, 
one can construct a factor graph by introducing factor nodes $\hi$'s for each local constraint $ \LocalConstraint{i}$, see \cref{fig:Graph}(b).
Even if the underlying graph $\Graph$ is tree-like as illustrated in \cref{fig:Graph}(a), the resulting factor graph contains many small loops as can be seen in \cref{fig:Graph}(b).
This is notoriously a problem if we want to implement a cavity equation naively since this method is guaranteed to work well only on tree-like factor graphs. \DIFdelbegin 

\DIFdelend One way to overcome this problem is to consider a particular form of Generalized Belief Propagation (GBP)  \cite{yedidia2001characterization} proposed precisely to deal with loops.
In the original formulation, the difficulty arising from the presence of loops is alleviated considering messages between regions of nodes that have a tree-like structure.
In the present case, this comes at the price of doubling the state space and considering the super-factor graph shown in \cref{fig:SuperFactor}.

In this representation we introduce a super variable for each link $(i,j)$ of $\Graph$ to hold two spin trajectories $S_{ij} = (\us_{ij},\us_{ji})$.
Then, this redundancy due to doubling the spin dimension is removed by adding delta function constraints to each factor node $\hi$ to ensure that all spin variables sharing the same first index $i$ are bound to be the same, i.e, $\us_{ij} = \us_{i}$ for all $j  \in \Neighbor{i}$.
From the above setting, we achieve a tree-like factor graph without losing the locality of constraints.
This trick has been recently used in similar contexts \cite{lokhov2015dynamic,rocchi2018slow}.

Having established a locally tree-like factor graph, it is straightforward to write the corresponding BP equations.
Following the standard procedure, the message that factor $\hat{i}$ sends to node $(ij)$ reads
\begin{equation}
\nu_{\hi \to (ij)} (\us_i, \us_j) \propto \sum_{ \ubs_\NeighborExcept{i}{j}} 
\LocalConstraint{i}
\prod_{k \in \NeighborExcept{i}{j}} \eta_{(ki)\to \hi} (\us_k , \us_i) 
\label{NuSFG}
\end{equation}
where the first summation runs over all possible sub-trajectories of spins $ \NeighborExcept{i}{j} $.
Similarly, the messages from the super-variables to the super-factors can be written trivially as there exist only two links for each super-variables, namely
\begin{equation}
\eta_{(ij) \to \hj} (\us_i, \us_j ) \propto \nu_{\hi \to (ij)} (\us_i, \us_j ).
\label{EtaSFG}
\end{equation}
By combining \cref{NuSFG,EtaSFG}, we can define a single set of messages $m_{i \to j} (\sigma_i, \sigma_j) \equiv \eta_{(ij) \to \hj} (S_{ij}) \propto \nu_{\hi \to (ij)} (S_{ij})$ satisfying \cref{Messageitoj}.

Once the messages are determined through the iterative process with \cref{Messageitoj} or equivalently \cref{NuSFG,EtaSFG}, 
the partition function \cref{PartitionFunctionDef} can be viewed as the sum of three different contributions coming from super-nodes,  super-factors, and super-links, namely, the partition function reads
\begin{equation}
\log Z_{L} = \sum_{(ij)} \log z_{(ij)} + \sum_{\hi} \log z_{\hi} - \sum_{<\hi,(ij)>} z_{\hi, (ij)}\:,
\label{PartitionLogFull}
\end{equation}
where the first sum is made on all super-nodes, the second one on all super-factors and the third one on all the links between super-nodes and super-factors.
For a general factor graph, these factors are given in terms of $\nu_{\hi \to (ij)}$ and $\eta_{(ij) \to \hi}$, i.e.,
\begin{equation}
z_{(ij)} = \sum_{S_{ij}} \nu_{\hi \to (ij)} (S_{ij}) \nu_{\hj \to (ij)} (S_{ji})\:,
\label{Znode}
\end{equation}
\begin{equation}
z_{\hi} = \sum_{\underline{S}_{\partial \hi}} \psi_{\hi}^S (\boldsymbol{S}_{\partial \hi})  \prod_{k \in \partial i } \eta_{(ki)\to \hi} (S_{ki})\:,
\label{Zfactor}
\end{equation}
\begin{equation}
z_{\hi, (ij)} = \sum_{S_{ij}} \nu_{\hi \to (ij)} (S_{ij}) \eta_{(ij) \to \hi} (S_{ij})\:,
\label{Zlink}
\end{equation}
where $\psi_{\hi}^S (\boldsymbol{S}_{\partial \hi})$ denotes the constraint on the super graph, 
\begin{equation}
\psi_{\hi}^S (\boldsymbol{S}_{\partial \hi}) = \LocalConstraint{i} \prod_{j \in \partial i } \delta(\underline{\sigma}_{ij} - \underline{\sigma}_{i} )
\end{equation}

In our setting, these equations are compactly written in terms of the messages \cref{Messageitoj}:
\begin{equation}
z_{(ij)} = \sum_{\sigma_i, \sigma_j} \Message{i}{j} \Message{j}{i}
\end{equation}
and
\begin{equation}
z_i = z_{\hi} = \sum_{\sigma_i} \sum_{\us_{\partial i}} \LocalConstraint{i} \prod_{k \in \partial i } \Message{k}{i}
\end{equation}
where we have introduced another representation $z_i = z_{\hi}$ to drop the hat from $\hi$, which is possible because factor nodes and original nodes are in one-to-one correspondence $\hi \leftrightarrow i$.
Additionally, one can immediately establish the relations $ z_{\hi,(ij)} = z_{\hj,(ji)} $ and $z_{(ij)}=z_{\hi,(ij)}$.
After making changes with these simplifications to \cref{PartitionLogFull}, one establishes \cref{PartitionFunctionFinal}.

\section{Motifs preventing cycles of lengths $L <4$}
\label{appendix:Motif}

\begin{figure}
	\centering
	\includegraphics[width=75mm]{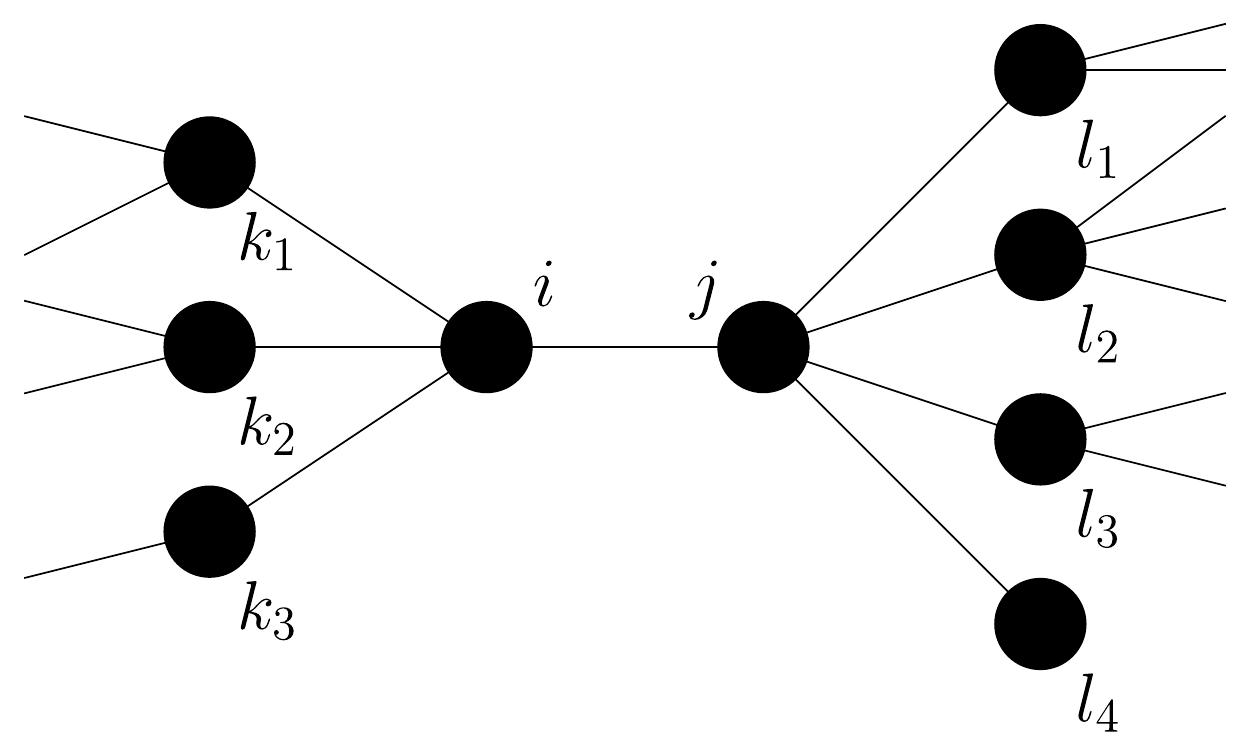}
	\caption{
		Illustration of graph motifs that forbid cycles with $L<4$. If the coupling matrix satisfies both conditions \cref{CouplingI,CouplingJ}, the states of $i$ and $j$ are completely independent of the rest of the system. At the same time, the states can only form a periodic orbit of length $4$, implying that there exist only cycles of length a multiple of $4$ in the system.
	}
	\label{fig:Motif}
\end{figure}

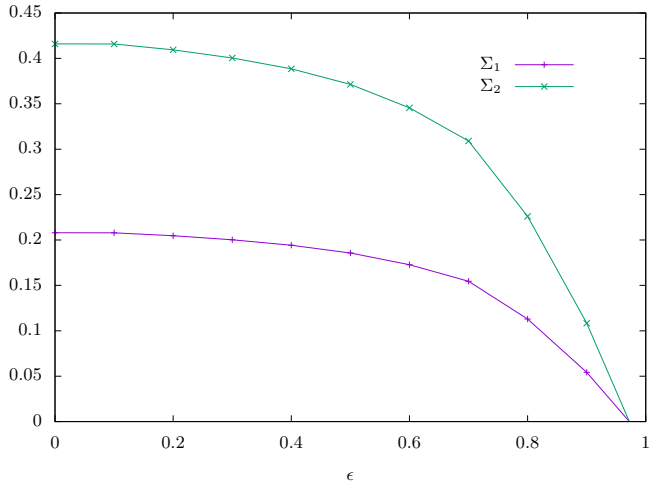
\begin{figure}
\centerline{\resizebox{1.1\linewidth}{!}{\input{Fig/fig9.tex}}}
\caption{Complexity of cycles of length $L=1$ and $L=2$ for a RR graph with $c=6$ as a function of $\epsilon$ for $N=100$. As mentioned in the text, these cycles disappears as $N$ increases. However, at finite $N$ and small $\epsilon$, the probability of the existence of the motif discussed in the text is small and numerically it is not found.}
\label{fig:fig-Comp-1-2}
\end{figure}

There exist certain motifs that forbid the existence of cycles of length $L<4$. 
Let us consider a subset of graph $\Graph$ given by \cref{fig:Motif} where two nodes $i$ and $j$ are connected.
Now, let us suppose that the couplings satisfy the following condition
\begin{align}
J_{ij} > \sum_{k \in \NeighborExcept{i}{j}} | J_{ik} |.
\label{CouplingI}
\end{align}
If this condition is met, the state of spin $\sigma_i^t$ is solely determined to be  $\sigma_j^{t-1}$ by the state of the spin $j$ at time step $(t-1)$ according to \cref{Dynamics}.
Additionally, let us impose a similar condition for the node $j$ but with the negative sign:
\begin{align}
- J_{ji} > \sum_{l \in \NeighborExcept{j}{i}} | J_{jl} |.
\label{CouplingJ}
\end{align}
Thus, it similarly follows that $ \sigma_j^{t}  = - \sigma_i^{t-1} $.  
The existence of such conditions implies that even though the network is connected, there can be subsets of a network that are independent of the rest of the network. 
In this particular case, because of the imposed conditions, the pair $(\sigma_i^t, \sigma_j^t)$ can only form a periodic orbit of length 4, thus the length of the cycles in this system should be a multiple of 4.
Considering that all the graph ensembles discussed in this paper have the finite probability per node of having such motifs as long as  $\epsilon>0$ and the connectivity is finite, the cycles of length shorter than $L<4$ cannot exist in the thermodynamic limit. This result becomes more and more likely as $\epsilon$ increases. In fact, for small values of $\epsilon$ the couplings tend to be more symmetric and the occurrences of this motif is suppressed. In this case, the existence of cycles of length smaller than $L=4$ is possible at finite sizes, as shown in Fig. \ref{fig:fig-Comp-1-2}.

\begin{figure}
\centering
\includegraphics[width=0.8\linewidth]{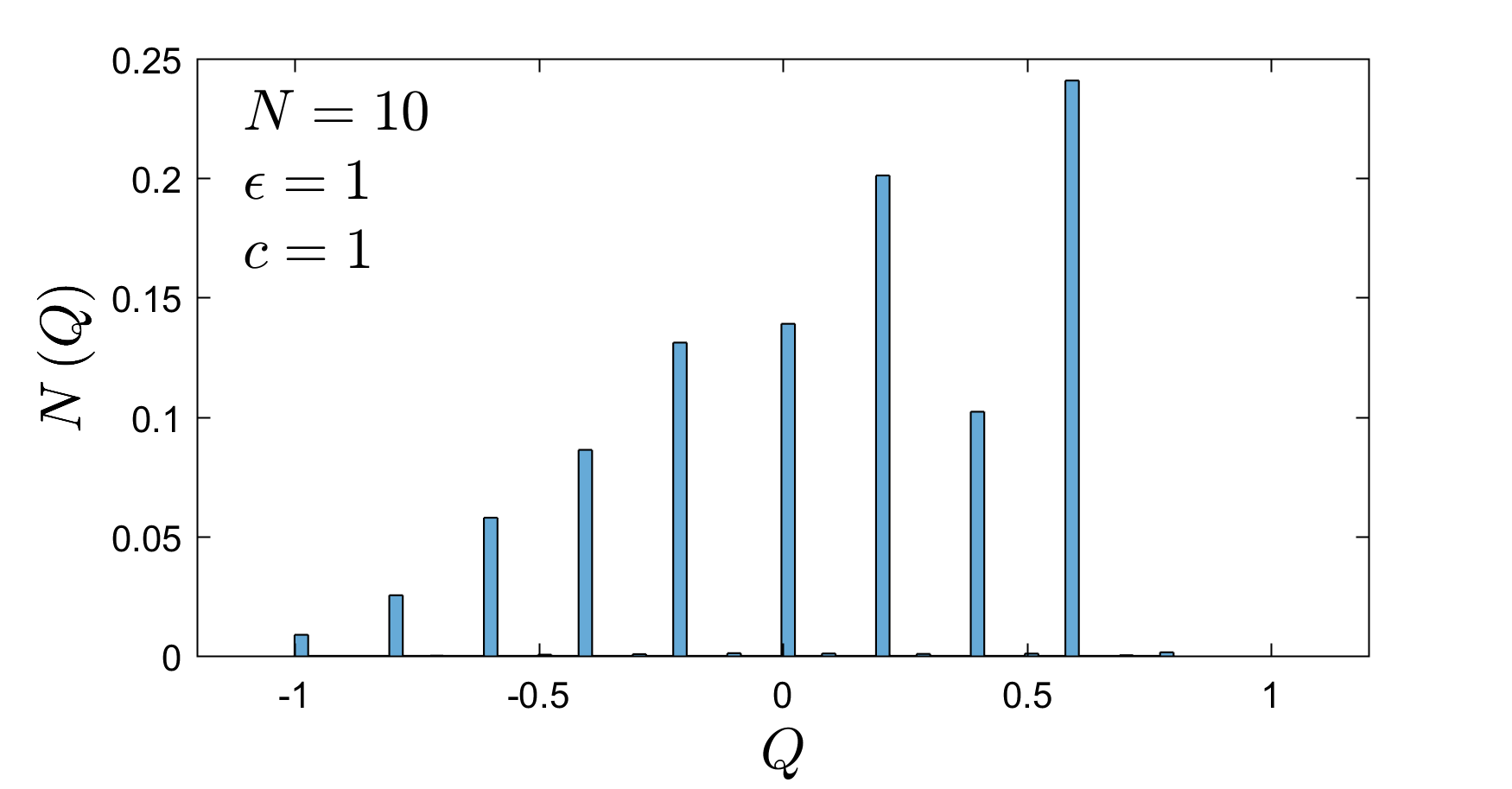}
\includegraphics[width=0.8\linewidth]{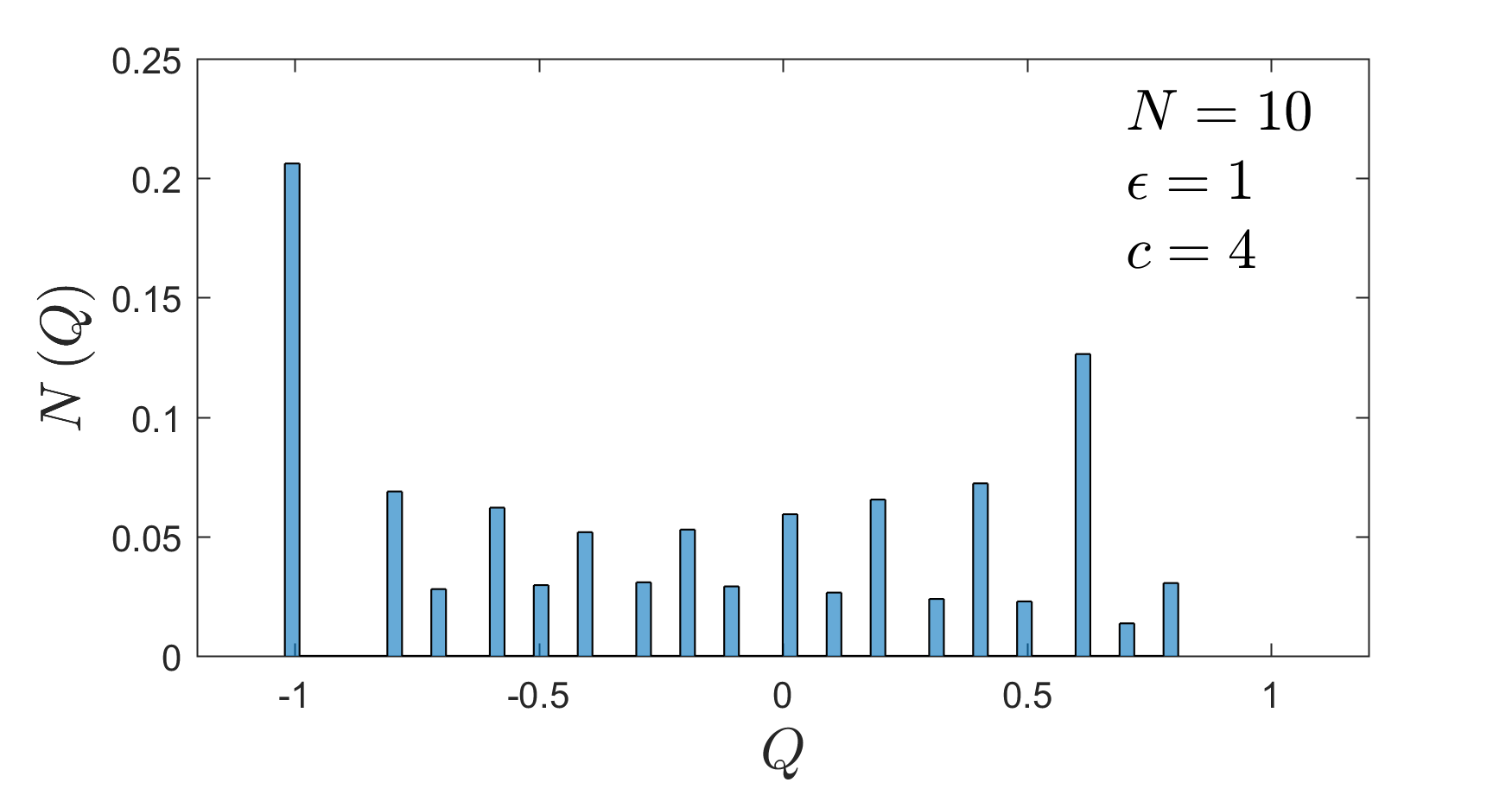}
\caption{
Histogram of the values of $Q$ for $N=10$, $\epsilon=1$ and connectivity $c=1$ (top) or $c=4$ (bottom) in the RR ensemble.
} 
\label{histC4}
\end{figure}

\begin{figure}
\centering
\includegraphics[width=0.75\linewidth]{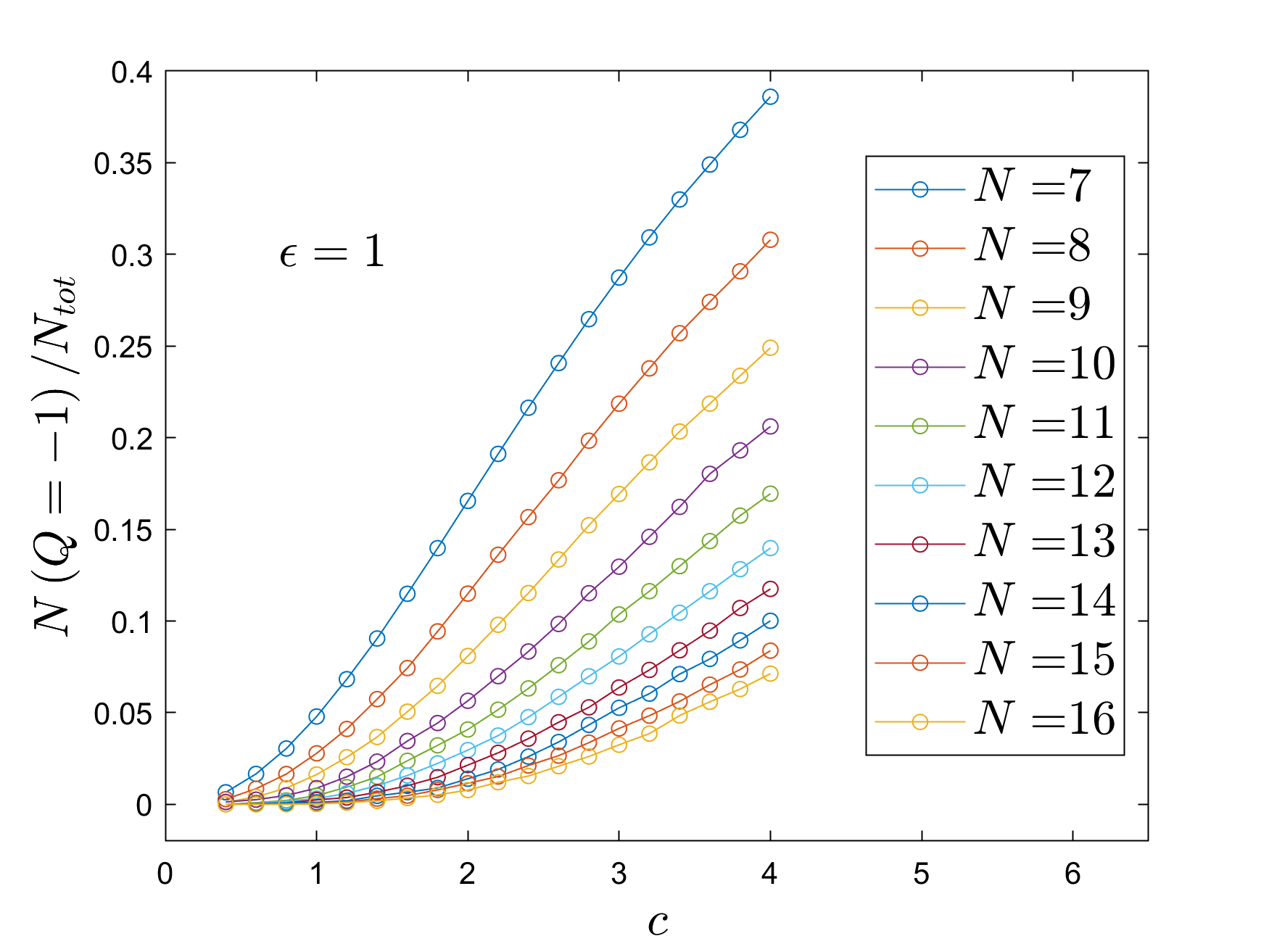}
\caption{Fraction of limit cycles of length 4 with $Q=-1$ as a function of the connectivity $c$ for various $N$, ranging from 7 to 16, at $\epsilon=1$  obtained for the RR ensemble.} \label{1meanC}
\end{figure}

If the two conditions defined in~\cref{CouplingI} and~\cref{CouplingJ} holds for each couple of link, cycles are skew symmetric. 
 A generic 4-cycle 
\[
\bs_1 \rightarrow \bs_2 \rightarrow \bs_3 \rightarrow \bs_4 \rightarrow \bs_1 \rightarrow \cdots
\]
is skew symmetric if 
\[
\bs_1 \rightarrow \bs_2 \rightarrow -\bs_1 \rightarrow -\bs_2 \rightarrow \bs_1 \rightarrow \cdots
\]
Such symmetry may be conveniently described by the quantity $Q$:
\begin{equation}
Q=\frac{1}{2N} \left(\bs_{1} \cdot \bs_{3}+ \bs_{2} \cdot \bs_{4} \right).
\end{equation} 
Because $\sigma = \{-1,1\}$, which means that $\vert \bs \vert^2=N$, we have $ -1 \le Q< 1$.
If a cycle is skew symmetric ($\sigma_1=-\sigma_3$ and $\sigma_2=-\sigma_4$), then $Q=-1$. Additionally, if we consider only limit cycles of length 4, $Q$ cannot be equal to 1. 
At $\epsilon=2$, coupling are perfectly antisymmetric and our numerical results confirm that the limit cycles are of length $L=4$, and all of them have $Q=-1$. We then ran simulations and checked the skew symmetry of limit cycles of length 4 to study the $Q$ values with decreasing connectivity parameter at $\epsilon$=1 (from $N=6$ to $N=15$ and $c$ ranging from 0.4 to 4).
As an example,  in~\cref{histC4}  we report two histograms of the obtained $Q$ values for $N=10$, in the case of $c=1$ and $c=4$ ($\epsilon=1$). It is evident that the fraction of limit cycles with $Q=-1$, which dominates in the case of $c=4$, tends to disappear when considering the case of $c=1$.
To confirm this data, Fig. \ref{1meanC} shows the fraction of limit cycles of length 4 for which $Q = -1$, as a function of $c$. In this case, all curves decrease on decreasing connectivity, consistently with the results of the previous figure. It is worth noticing that for large $N$ the occurrences of situations like the one outlined in \cref{fig:Motif}, which can be found only at sufficiently small values of $c$, increase, as well as the probability that $\sign(J_{ij})=\sign(J_{ji})$ (given that $\epsilon=1$) in at least one pair of nodes, which is sufficient to break the skew-symmetry of the limit cycles of length 4. The second effect dominates over the first one at low connectivities and skew symmetricity disappears. 

\section{Additivity of complexity in the presence of disconnected clusters}
%

In this appendix, we construct a simple relation for $Z_L$ for the case of graphs with a finite number of connected clusters. 
Since each connected cluster is independent of the other ones, we have the following additivity property:
\begin{align}
\ln Z_L = N_C \sum_C P(C) \ln Z_C, 
\end{align}
where $N_C$ is the number of connected clusters and $P(C)$ is the cluster size distribution.
From the normalization condition, we have the following relation
\begin{align}
	N_C \sum_C P(C) C = N.
\end{align} 
If the clusters possess the same statistical properties and their sizes are sufficiently large, it is reasonable to assume that $\ln Z_C$  follows the same asymptotic expansion of $\ln Z_L$ i.e., $ \ln Z_C \sim C \Sigma_{L}  + B_L + O(C^{-1})$.
Under these conditions, the overall partition function reads
\begin{align}
\ln Z_L & \sim  N \Sigma_L + B_L + O(N^{-1}) \nonumber \\
& = N_C \sum_C P(C) \ln Z_C  \nonumber  \\
& \sim N \Sigma_L + N_C B_L + O(N_C C^{-1}),
\label{TreeRelation}
\end{align}
where we have used the normalization condition.
Additionally, using eq. (\ref{PartitionFunctionQuenched}), we reach a mismatch in the constant order, thus implying $B_L =0$. 
This relation can potentially become wrong if $N_C$ is extensive. In this case all the higher corrections in \cref{TreeRelation} become of the same order as $\Sigma_L$. 
Nevertheless, we found numerically that $B_L$ is distributed close to our prediction $B_L =0$.

\FloatBarrier

\bibliographystyle{unsrt}
\bibliography{ref}

\end{document}

%% file: Fig/fig5.tex
\begingroup
  \makeatletter
  \providecommand\color[2][]{%
    \GenericError{(gnuplot) \space\space\space\@spaces}{%
      Package color not loaded in conjunction with
      terminal option `colourtext'%
    }{See the gnuplot documentation for explanation.%
    }{Either use 'blacktext' in gnuplot or load the package
      color.sty in LaTeX.}%
    \renewcommand\color[2][]{}%
  }%
  \providecommand\includegraphics[2][]{%
    \GenericError{(gnuplot) \space\space\space\@spaces}{%
      Package graphicx or graphics not loaded%
    }{See the gnuplot documentation for explanation.%
    }{The gnuplot epslatex terminal needs graphicx.sty or graphics.sty.}%
    \renewcommand\includegraphics[2][]{}%
  }%
  \providecommand\rotatebox[2]{#2}%
  \@ifundefined{ifGPcolor}{%
    \newif\ifGPcolor
    \GPcolortrue
  }{}%
  \@ifundefined{ifGPblacktext}{%
    \newif\ifGPblacktext
    \GPblacktexttrue
  }{}%
  \let\gplgaddtomacro\g@addto@macro
  \gdef\gplbacktext{}%
  \gdef\gplfronttext{}%
  \makeatother
  \ifGPblacktext
    \def\colorrgb#1{}%
    \def\colorgray#1{}%
  \else
    \ifGPcolor
      \def\colorrgb#1{\color[rgb]{#1}}%
      \def\colorgray#1{\color[gray]{#1}}%
      \expandafter\def\csname LTw\endcsname{\color{white}}%
      \expandafter\def\csname LTb\endcsname{\color{black}}%
      \expandafter\def\csname LTa\endcsname{\color{black}}%
      \expandafter\def\csname LT0\endcsname{\color[rgb]{1,0,0}}%
      \expandafter\def\csname LT1\endcsname{\color[rgb]{0,1,0}}%
      \expandafter\def\csname LT2\endcsname{\color[rgb]{0,0,1}}%
      \expandafter\def\csname LT3\endcsname{\color[rgb]{1,0,1}}%
      \expandafter\def\csname LT4\endcsname{\color[rgb]{0,1,1}}%
      \expandafter\def\csname LT5\endcsname{\color[rgb]{1,1,0}}%
      \expandafter\def\csname LT6\endcsname{\color[rgb]{0,0,0}}%
      \expandafter\def\csname LT7\endcsname{\color[rgb]{1,0.3,0}}%
      \expandafter\def\csname LT8\endcsname{\color[rgb]{0.5,0.5,0.5}}%
    \else
      \def\colorrgb#1{\color{black}}%
      \def\colorgray#1{\color[gray]{#1}}%
      \expandafter\def\csname LTw\endcsname{\color{white}}%
      \expandafter\def\csname LTb\endcsname{\color{black}}%
      \expandafter\def\csname LTa\endcsname{\color{black}}%
      \expandafter\def\csname LT0\endcsname{\color{black}}%
      \expandafter\def\csname LT1\endcsname{\color{black}}%
      \expandafter\def\csname LT2\endcsname{\color{black}}%
      \expandafter\def\csname LT3\endcsname{\color{black}}%
      \expandafter\def\csname LT4\endcsname{\color{black}}%
      \expandafter\def\csname LT5\endcsname{\color{black}}%
      \expandafter\def\csname LT6\endcsname{\color{black}}%
      \expandafter\def\csname LT7\endcsname{\color{black}}%
      \expandafter\def\csname LT8\endcsname{\color{black}}%
    \fi
  \fi
    \setlength{\unitlength}{0.0500bp}%
    \ifx\gptboxheight\undefined%
      \newlength{\gptboxheight}%
      \newlength{\gptboxwidth}%
      \newsavebox{\gptboxtext}%
    \fi%
    \setlength{\fboxrule}{0.5pt}%
    \setlength{\fboxsep}{1pt}%
\begin{picture}(7200.00,5040.00)%
    \gplgaddtomacro\gplbacktext{%
      \csname LTb\endcsname
      \put(814,704){\makebox(0,0)[r]{\strut{}$0$}}%
      \put(814,1527){\makebox(0,0)[r]{\strut{}$0.2$}}%
      \put(814,2350){\makebox(0,0)[r]{\strut{}$0.4$}}%
      \put(814,3173){\makebox(0,0)[r]{\strut{}$0.6$}}%
      \put(814,3996){\makebox(0,0)[r]{\strut{}$0.8$}}%
      \put(814,4819){\makebox(0,0)[r]{\strut{}$1$}}%
      \put(946,484){\makebox(0,0){\strut{}$0$}}%
      \put(1678,484){\makebox(0,0){\strut{}$0.5$}}%
      \put(2410,484){\makebox(0,0){\strut{}$1$}}%
      \put(3142,484){\makebox(0,0){\strut{}$1.5$}}%
      \put(3875,484){\makebox(0,0){\strut{}$2$}}%
      \put(4607,484){\makebox(0,0){\strut{}$2.5$}}%
      \put(5339,484){\makebox(0,0){\strut{}$3$}}%
      \put(6071,484){\makebox(0,0){\strut{}$3.5$}}%
      \put(6803,484){\makebox(0,0){\strut{}$4$}}%
    }%
    \gplgaddtomacro\gplfronttext{%
      \csname LTb\endcsname
      \put(198,2761){\rotatebox{-270}{\makebox(0,0){\strut{}$\Sigma_4$}}}%
      \put(3874,154){\makebox(0,0){\strut{}$c$}}%
      \csname LTb\endcsname
      \put(1555,4298){\makebox(0,0)[r]{\strut{}DP}}%
      \csname LTb\endcsname
      \put(1555,4078){\makebox(0,0)[r]{\strut{}ER}}%
      \csname LTb\endcsname
      \put(1555,3858){\makebox(0,0)[r]{\strut{}RR}}%
    }%
    \gplgaddtomacro\gplbacktext{%
      \csname LTb\endcsname
      \put(3478,2820){\makebox(0,0)[r]{\strut{}$0$}}%
      \put(3478,3139){\makebox(0,0)[r]{\strut{}$0.2$}}%
      \put(3478,3458){\makebox(0,0)[r]{\strut{}$0.4$}}%
      \put(3478,3778){\makebox(0,0)[r]{\strut{}$0.6$}}%
      \put(3478,4097){\makebox(0,0)[r]{\strut{}$0.8$}}%
      \put(3478,4416){\makebox(0,0)[r]{\strut{}$1$}}%
      \put(3610,2600){\makebox(0,0){\strut{}$0$}}%
      \put(3982,2600){\makebox(0,0){\strut{}$0.5$}}%
      \put(4354,2600){\makebox(0,0){\strut{}$1$}}%
      \put(4726,2600){\makebox(0,0){\strut{}$1.5$}}%
      \put(5099,2600){\makebox(0,0){\strut{}$2$}}%
      \put(5471,2600){\makebox(0,0){\strut{}$2.5$}}%
      \put(5843,2600){\makebox(0,0){\strut{}$3$}}%
      \put(6215,2600){\makebox(0,0){\strut{}$3.5$}}%
      \put(6587,2600){\makebox(0,0){\strut{}$4$}}%
    }%
    \gplgaddtomacro\gplfronttext{%
      \csname LTb\endcsname
      \put(2862,3618){\rotatebox{-270}{\makebox(0,0){\strut{}$|C|_{\text{MAX}}$}}}%
      \put(5098,2270){\makebox(0,0){\strut{}$c$}}%
      \csname LTb\endcsname
      \put(5434,3668){\makebox(0,0)[r]{\strut{}DP}}%
      \csname LTb\endcsname
      \put(5434,3448){\makebox(0,0)[r]{\strut{}ER}}%
      \csname LTb\endcsname
      \put(5434,3228){\makebox(0,0)[r]{\strut{}RR}}%
    }%
    \gplbacktext
    \put(0,0){\includegraphics{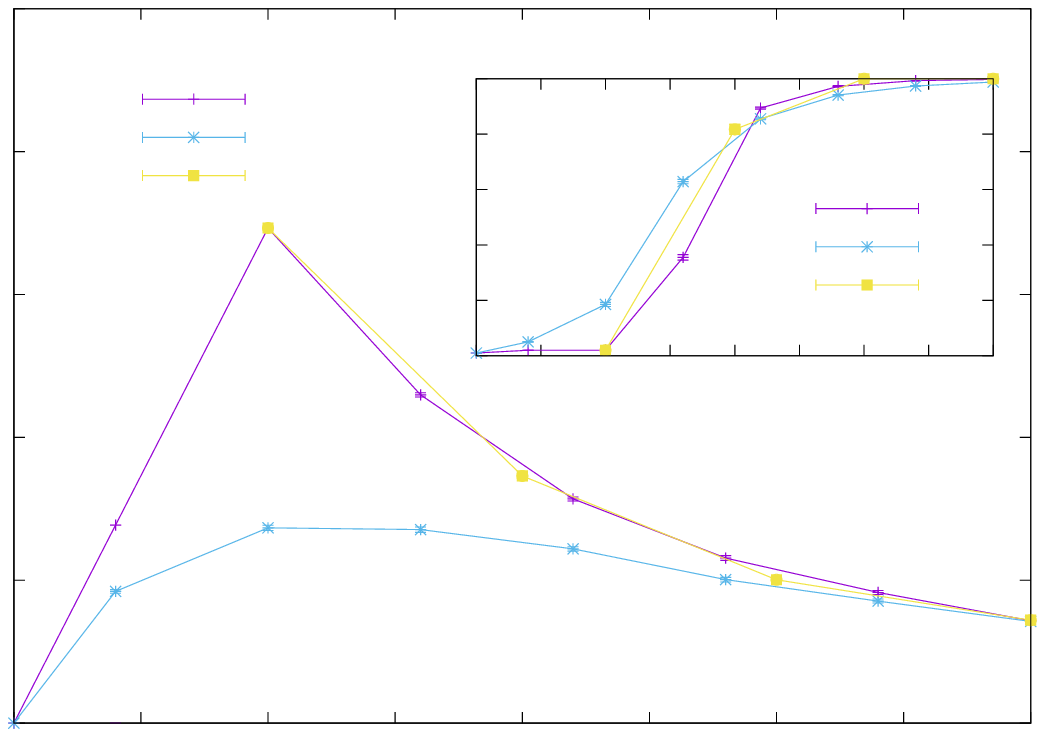}}%
    \gplfronttext
  \end{picture}%
\endgroup

%% file: Fig/fig9.tex
\begingroup
  \makeatletter
  \providecommand\color[2][]{%
    \GenericError{(gnuplot) \space\space\space\@spaces}{%
      Package color not loaded in conjunction with
      terminal option `colourtext'%
    }{See the gnuplot documentation for explanation.%
    }{Either use 'blacktext' in gnuplot or load the package
      color.sty in LaTeX.}%
    \renewcommand\color[2][]{}%
  }%
  \providecommand\includegraphics[2][]{%
    \GenericError{(gnuplot) \space\space\space\@spaces}{%
      Package graphicx or graphics not loaded%
    }{See the gnuplot documentation for explanation.%
    }{The gnuplot epslatex terminal needs graphicx.sty or graphics.sty.}%
    \renewcommand\includegraphics[2][]{}%
  }%
  \providecommand\rotatebox[2]{#2}%
  \@ifundefined{ifGPcolor}{%
    \newif\ifGPcolor
    \GPcolortrue
  }{}%
  \@ifundefined{ifGPblacktext}{%
    \newif\ifGPblacktext
    \GPblacktexttrue
  }{}%
  \let\gplgaddtomacro\g@addto@macro
  \gdef\gplbacktext{}%
  \gdef\gplfronttext{}%
  \makeatother
  \ifGPblacktext
    \def\colorrgb#1{}%
    \def\colorgray#1{}%
  \else
    \ifGPcolor
      \def\colorrgb#1{\color[rgb]{#1}}%
      \def\colorgray#1{\color[gray]{#1}}%
      \expandafter\def\csname LTw\endcsname{\color{white}}%
      \expandafter\def\csname LTb\endcsname{\color{black}}%
      \expandafter\def\csname LTa\endcsname{\color{black}}%
      \expandafter\def\csname LT0\endcsname{\color[rgb]{1,0,0}}%
      \expandafter\def\csname LT1\endcsname{\color[rgb]{0,1,0}}%
      \expandafter\def\csname LT2\endcsname{\color[rgb]{0,0,1}}%
      \expandafter\def\csname LT3\endcsname{\color[rgb]{1,0,1}}%
      \expandafter\def\csname LT4\endcsname{\color[rgb]{0,1,1}}%
      \expandafter\def\csname LT5\endcsname{\color[rgb]{1,1,0}}%
      \expandafter\def\csname LT6\endcsname{\color[rgb]{0,0,0}}%
      \expandafter\def\csname LT7\endcsname{\color[rgb]{1,0.3,0}}%
      \expandafter\def\csname LT8\endcsname{\color[rgb]{0.5,0.5,0.5}}%
    \else
      \def\colorrgb#1{\color{black}}%
      \def\colorgray#1{\color[gray]{#1}}%
      \expandafter\def\csname LTw\endcsname{\color{white}}%
      \expandafter\def\csname LTb\endcsname{\color{black}}%
      \expandafter\def\csname LTa\endcsname{\color{black}}%
      \expandafter\def\csname LT0\endcsname{\color{black}}%
      \expandafter\def\csname LT1\endcsname{\color{black}}%
      \expandafter\def\csname LT2\endcsname{\color{black}}%
      \expandafter\def\csname LT3\endcsname{\color{black}}%
      \expandafter\def\csname LT4\endcsname{\color{black}}%
      \expandafter\def\csname LT5\endcsname{\color{black}}%
      \expandafter\def\csname LT6\endcsname{\color{black}}%
      \expandafter\def\csname LT7\endcsname{\color{black}}%
      \expandafter\def\csname LT8\endcsname{\color{black}}%
    \fi
  \fi
    \setlength{\unitlength}{0.0500bp}%
    \ifx\gptboxheight\undefined%
      \newlength{\gptboxheight}%
      \newlength{\gptboxwidth}%
      \newsavebox{\gptboxtext}%
    \fi%
    \setlength{\fboxrule}{0.5pt}%
    \setlength{\fboxsep}{1pt}%
\begin{picture}(7200.00,5040.00)%
    \gplgaddtomacro\gplbacktext{%
      \csname LTb\endcsname
      \put(726,704){\makebox(0,0)[r]{\strut{}$0$}}%
      \put(726,1161){\makebox(0,0)[r]{\strut{}$0.05$}}%
      \put(726,1618){\makebox(0,0)[r]{\strut{}$0.1$}}%
      \put(726,2076){\makebox(0,0)[r]{\strut{}$0.15$}}%
      \put(726,2533){\makebox(0,0)[r]{\strut{}$0.2$}}%
      \put(726,2990){\makebox(0,0)[r]{\strut{}$0.25$}}%
      \put(726,3447){\makebox(0,0)[r]{\strut{}$0.3$}}%
      \put(726,3905){\makebox(0,0)[r]{\strut{}$0.35$}}%
      \put(726,4362){\makebox(0,0)[r]{\strut{}$0.4$}}%
      \put(726,4819){\makebox(0,0)[r]{\strut{}$0.45$}}%
      \put(858,484){\makebox(0,0){\strut{}$0$}}%
      \put(2047,484){\makebox(0,0){\strut{}$0.2$}}%
      \put(3236,484){\makebox(0,0){\strut{}$0.4$}}%
      \put(4425,484){\makebox(0,0){\strut{}$0.6$}}%
      \put(5614,484){\makebox(0,0){\strut{}$0.8$}}%
      \put(6803,484){\makebox(0,0){\strut{}$1$}}%
    }%
    \gplgaddtomacro\gplfronttext{%
      \csname LTb\endcsname
      \put(3830,154){\makebox(0,0){\strut{}$\epsilon$}}%
      \csname LTb\endcsname
      \put(5354,4298){\makebox(0,0)[r]{\strut{}$\Sigma_1$}}%
      \csname LTb\endcsname
      \put(5354,4078){\makebox(0,0)[r]{\strut{}$\Sigma_2$}}%
    }%
    \gplbacktext
    \put(0,0){\includegraphics{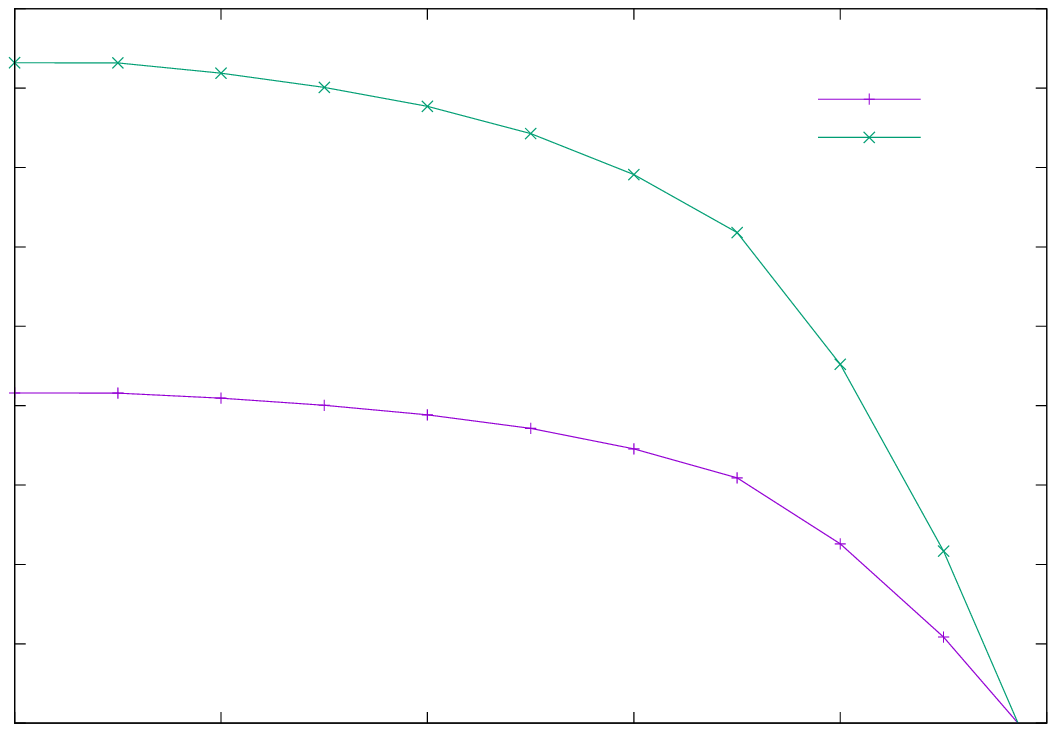}}%
    \gplfronttext
  \end{picture}%
\endgroup